\documentclass[a4paper,11pt]{article}
\pdfoutput=1 
\usepackage{jheppub} 
\usepackage[T1]{fontenc} 

\usepackage{epsfig}
\usepackage{graphicx}
\usepackage{dcolumn}
\usepackage{bm}
\usepackage{ltablex,booktabs}
\usepackage{overpic}
\usepackage{subfigure}
\usepackage{float}
\usepackage{color}
\usepackage{amsmath}
\usepackage{mathcomp}
\usepackage{mathrsfs}
\usepackage{multirow}
\usepackage{rotating}
\usepackage{amssymb}
\usepackage{gensymb}
\usepackage{amsmath}
\usepackage{tabularx}
\usepackage{epstopdf}

\title{Amplitude analysis and branching fraction measurement of the decay \boldmath $D_{s}^{+} \to \pi^{+}\pi^{0}\pi^{0}$}
\collaborationImg{\includegraphics[width=0.15\textwidth, angle=90]{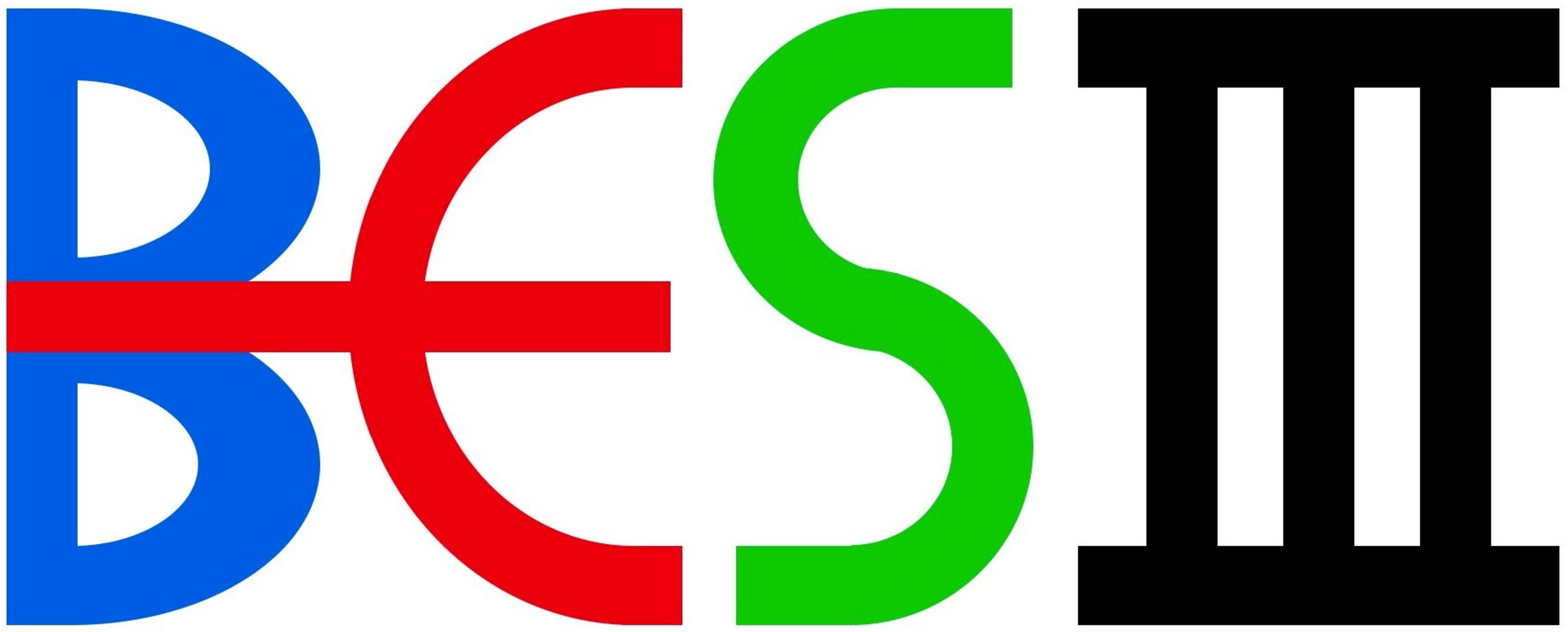}}
\collaboration{BESIII Collaboration}

\author{
M.~Ablikim$^{1}$, M.~N.~Achasov$^{10,b}$, P.~Adlarson$^{66}$, S. ~Ahmed$^{14}$, M.~Albrecht$^{4}$, R.~Aliberti$^{27}$, A.~Amoroso$^{65A,65C}$, M.~R.~An$^{31}$, Q.~An$^{62,48}$, X.~H.~Bai$^{56}$
% ...
\\
\vspace{0.2cm} {\it
$^{1}$ Institute of High Energy Physics, Beijing 100049, People's Republic of China\\
% ...
$^{4}$ Bochum Ruhr-University, D-44780 Bochum, Germany\\
% ...
$^{10}$ G.I. Budker Institute of Nuclear Physics SB RAS (BINP), Novosibirsk 630090, Russia\\
% ...
$^{14}$ Helmholtz Institute Mainz, Staudinger Weg 18, D-55099 Mainz, Germany\\
% ...
$^{27}$ Johannes Gutenberg University of Mainz, Johann-Joachim-Becher-Weg 45, D-55099 Mainz, Germany\\
% ...
$^{31}$ Liaoning Normal University, Dalian 116029, People's Republic of China\\
% ...
$^{48}$ State Key Laboratory of Particle Detection and Electronics, Beijing 100049, Hefei 230026, People's Republic of China\\
% ...
$^{56}$ University of Jinan, Jinan 250022, People's Republic of China\\
% ...
$^{62}$ University of Science and Technology of China, Hefei 230026, People's Republic of China\\
% ...
$^{65}$ University of Turin and INFN, (A)University of Turin, I-10125, Turin, Italy; (B)University of Eastern Piedmont, I-15121, Alessandria, Italy; (C)INFN, I-10125, Turin, Italy\\
$^{66}$ Uppsala University, Box 516, SE-75120 Uppsala, Sweden\\
% ...
$^{b}$ Also at the Novosibirsk State University, Novosibirsk, 630090, Russia\\
% ...
}
}
\date{\today}
\abstract{
  Using a data set corresponding to an integrated luminosity of 6.32~$\rm fb^{-1}$
  recorded by the BESIII detector at center-of-mass energies between 4.178 and
  4.226~GeV, an amplitude analysis of the decay
  $D_{s}^{+} \to \pi^{+}\pi^{0}\pi^{0}$ is performed, and the relative fractions
  and phases of different intermediate processes are determined.
  The absolute branching fraction of the decay $D_{s}^{+} \to \pi^{+}\pi^{0}\pi^{0}$ is
  measured to be $(0.50\pm 0.04_{\text{stat}}\pm 0.02_{\text{syst}})\%$.
  The absolute branching fraction of the intermediate process
  $D_{s}^{+} \to f_0(980)\pi^{+}, f_0(980)\to\pi^{0}\pi^{0}$ is determined to
  be $(0.21\pm 0.03_{\text{stat}}\pm 0.03_{\text{syst}})\%$.
}

\arxivnumber{}
\begin{document}
\maketitle
\flushbottom


\section{Introduction}
The constituent quark model has been very successful in explaining the composition
of hadrons in the past few decades. In this model, the observed meson spectrum
is described as bound $q\bar{q}$ states grouped into SU(n) flavor multiplets.
The nonets of pseudo-scalar, vector and tensor mesons have been well
identified. Nevertheless, the identification of the scalar-meson nonet is still
ambiguous. Distinguishing scalar mesons from non-resonant background is
rather difficult due to their broad widths and non-distinctive angular
distribution. There are copious candidates for the $J^{PC} = 0^{++}$
nonets~\cite{PDG}. The case with isospin zero states, e.g.~$f_0(500)$,
$f_0(980)$, $f_0(1370)$, $f_0(1500)$, and $f_0(1710)$, is the most complicated from
both experimental and theoretical points of view. Among them, the $f_0(980)$
meson, as a possible tetraquark candidate~\cite{PhysRevD.27.588, PhysRevD.41.2236, PhysRevD.96.033002}, 
is particularly interesting and can be studied via the hadronic decays
$D_{s}^{+}\to\pi^{+}\pi^{0}\pi^{0}$, $D_{s}^{+} \to \pi^{+}\pi^{+}\pi^{-}$ and $D_{s}^{+} \to K^{+}K^{-}\pi^{+}$.
Charge conjugation is implied throughout in this paper.
The current published branching fraction (BF)  of $D^+_s\to f_{0(2)}\pi^+$
from the $D_{s}^{+} \to \pi^{+}\pi^{+}\pi^{-}$ decays has large discrepancies~\cite{PDG, E687:1997jvh, E791:2000lzz}
with that measured from the $D_{s}^{+} \to K^{+}K^{-}\pi^{+}$ decays. 
The $f_{0(2)}$ contributions may suffer 
from the contaminations of $a_{0}(980)\to K^+K^-$ or $\rho \to \pi^+ \pi^-$ and 
the $D_{s}^{+} \to \pi^{+}\pi^{0}\pi^{0}$ decays offer a cleaner
environment due to absence of these contributions. 

Furthermore, hadronic $D_{s}^{+}$
decays can be used to probe the interplay of short-distance weak-decay matrix
elements and long-distance QCD interactions, and the measured BFs provide valuable information concerning the amplitudes and
phases that induce in decay processes~\cite{Li2021iwf, BCKa0, PRD79-034016, PRD81-074021, PRD84-074019}.

The CLEO Collaboration reported a measurement of absolute BF
$\mathcal{B}(D_{s}^{+} \to \pi^{+}\pi^{0}\pi^{0}) = (0.65\pm 0.13)\%$~\cite{CLEO-BF},
using 600~pb$^{-1}$ of $e^+e^-$ collision data recorded at a center-of-mass
energy~($\sqrt{s}$) of 4.17~GeV. 
In this analysis, by using 6.32~$\rm fb^{-1}$ of data collected
with the BESIII detector ranging from $\sqrt{s}=4.178$~GeV to $\sqrt{s}=4.226$~GeV,
we perform the first
amplitude analysis of $D^+_s\to \pi^+\pi^0\pi^0$ and a more precise measurement
of its absolute BF. The amplitude analysis allows the determination of
$\mathcal{B}(D_{s}^{+} \to f_{0}(980)\pi^{+})$,
$\mathcal{B}(D_{s}^{+} \to f_{0}(1370)\pi^{+})$, and
$\mathcal{B}(D_{s}^{+} \to f_{2}(1270)\pi^{+})$.

\section{Detector and data sets}
\label{sec:detector_dataset}
The BESIII detector~\cite{Ablikim:2009aa} records symmetric $e^+e^-$ collisions
provided by the BEPCII storage ring in the range from $\sqrt{s}=2.00$~GeV~ to
$\sqrt{s}=4.95$~GeV~\cite{Yu:IPAC2016-TUYA01, Ablikim:2019hff}.
The cylindrical core
of the BESIII detector covers 93\% of the full solid angle and consists of a
helium-based multilayer drift chamber~(MDC), a plastic scintillator
time-of-flight system~(TOF), and a CsI(Tl) electromagnetic calorimeter~(EMC),
which are all enclosed in a superconducting solenoidal magnet providing 
a magnetic field of 1.0~T. The solenoid is supported by an octagonal flux-return yoke with
resistive plate counter muon identification modules interleaved with steel.
The charged-particle momentum resolution at $1~{\rm GeV}/c$ is $0.5\%$, and the
d$E$/d$x$ resolution is $6\%$ for electrons from Bhabha scattering. The EMC
measures photon energies with a resolution of $2.5\%$ ($5\%$) at $1$~GeV in the
barrel (end cap) region. The time resolution in the TOF barrel region is 68~ps,
while that in the end cap region is 110~ps. The end cap TOF
system was upgraded in 2015 using multi-gap resistive plate chamber
technology, providing a time resolution of 60~ps~\cite{etof1, etof2, etof3}.

The data samples used in this analysis are listed in Table~\ref{energe}~\cite{LumE42301,LumE42302}.
Since the cross section of $D_{s}^{*\pm}D_{s}^{\mp}$ production in $e^{+}e^{-}$
annihilation is about a factor of twenty larger than that of
$D_{s}^{+}D_{s}^{-}$~\cite{DsStrDs}, and the $D_{s}^{*+}$ meson decays to
$\gamma D_{s}^{+}$ with a dominant BF of $(93.5\pm0.7)$\%~\cite{PDG}, the
signal events discussed in this paper are selected from the process
$e^+e^-\to D_{s}^{*\pm}D_{s}^{\mp} \to \gamma D_{s}^{+}D_{s}^{-}$.

 \begin{table}[htb]
 \renewcommand\arraystretch{1.25}
 \centering
 \caption{The integrated luminosities ($\mathcal{L}_{\rm int}$) and the 
   requirements on $M_{\rm rec}$ for various collision energies. The
   definition of $M_{\rm rec}$ is given in Eq.~(\ref{eq:mrec}). The first and
   the second uncertainties are statistical and systematic, respectively.}
 \begin{tabular}{ccc}
 \hline
 $\sqrt{s}$ (GeV) & $\mathcal{L}_{\rm int}$ (pb$^{-1}$) & $M_{\rm rec}$ (GeV/$c^2$)\\
 \hline
  4.178 &3189.0$\pm$0.2$\pm$31.9&[2.050, 2.180] \\
  4.189 &526.7$\pm$0.1$\pm$2.2&[2.048, 2.190] \\
  4.199 &526.0$\pm$0.1$\pm$2.1&[2.046, 2.200] \\
  4.209 &517.1$\pm$0.1$\pm$1.8&[2.044, 2.210] \\
  4.219 &514.6$\pm$0.1$\pm$1.8&[2.042, 2.220] \\
  4.226 &1047.3$\pm$0.1$\pm10.2$&[2.040, 2.220] \\
  \hline
 \end{tabular}
 \label{energe}
\end{table}

Simulated data samples produced with a {\sc geant4}-based~\cite{GEANT4} Monte
Carlo (MC) package, which includes the geometric description of the BESIII
detector and the detector response, are used to determine detection
efficiencies and to estimate backgrounds. The simulation models the beam energy
spread and initial state radiation (ISR) in the $e^+e^-$ annihilations with the
generator {\sc kkmc}~\cite{KKMC1, KKMC2}. The inclusive MC sample includes the
production of open charm processes, the ISR production of vector
charmonium(-like) states, and the continuum processes incorporated in
{\sc kkmc}~\cite{KKMC1, KKMC2}. The known decay modes are modelled with
{\sc evtgen}~\cite{EVTGEN1, EVTGEN2} using BFs taken from the Particle Data
Group~\cite{PDG}, and the remaining unknown charmonium decays are modelled with
{\sc lundcharm}~\cite{LUNDCHARM1, LUNDCHARM2}. Final state radiation~(FSR) from
charged final state particles is incorporated using {\sc photos}~\cite{PHOTOS}.

\section{Event selection}
\label{ST-selection}

The data samples were collected just above the $D_s^{*\pm}D_s^{\mp}$
threshold, which allows to extract relatively pure samples for amplitude analysis and measurements of
absolute BFs of the hadronic $D^+_s$ meson decays with a tag method. 
The tag method has single-tag~(ST) and double-tag~(DT) candidates.
The ST candidates are those $D_{s}^{\pm}$ mesons without further
requirements on the remaining tracks and EMC showers. The DT
candidates are identified by fully reconstructing the
$D_s^+D_s^-$ mesons, where one of the $D_s^{\pm}$ mesons decays into the signal mode $D_{s}^{+} \to \pi^{+}\pi^{0}\pi^{0}$ and the
other to a tag mode. 
The $D_s^\pm$ mesons are reconstructed through the final state particles, i.e.~$\pi^\pm$, $K^\pm$,
$\eta$, $\eta^{\prime}$, $K_{S}^{0}$ and $\pi^{0}$, whose selection criteria is discussed below.

For charged tracks not originating from $K_S^0$ decays, the distance of closest
approach to the interaction point is required to be less than 10~cm along the
beam direction and less than 1~cm in the plane perpendicular to the beam.
Particle identification~(PID) for charged tracks combines measurements of the
specific ionization energy losses in the MDC~(d$E$/d$x$) and the flight time in
the TOF to form a likelihood $\mathcal{L}(h)~(h=K,\pi)$ for the
hypothesis of being a hadron $h$. A charged hadron is identified as a kaon if $\mathcal{L}(K)$ is larger than $\mathcal{L}(\pi)$, otherwise
it is identified as a pion.

The $K_{S}^0$ candidates are reconstructed from two oppositely charged tracks
satisfying $|\!\cos\theta|< 0.93$ and the distance of closest approach along
the beam direction must be less than 20~cm. The two charged tracks coming form
the $K^0_S$ are assigned as
$\pi^+\pi^-$ without imposing further PID criteria. They are constrained to
originate from a common vertex and are required to have an invariant mass
within $|M_{\pi^{+}\pi^{-}} - m_{K_{S}^{0}}|<$ 12~MeV$/c^{2}$, where
$m_{K_{S}^{0}}$ is the $K^0_{S}$ mass taken from PDG~\cite{PDG}. 

Photon candidates are identified using showers in the EMC. The deposited
energy of each shower must be more than 25~MeV in the barrel
region~($|\!\cos \theta|< 0.80$) and more than 50~MeV in the end cap
region~($0.86 <|\!\cos \theta|< 0.92$). The angle between the position of each
shower in the EMC and any charged track must be greater
than 10 degrees to exclude showers originating from charged tracks. The
difference between the EMC time and the event start time is required to be
within [0, 700]\,ns to suppress electronic noise and showers unrelated to the
event.

The $\pi^0$ $(\eta)$ candidates are reconstructed through
$\pi^0\to \gamma\gamma$ ($\eta \to \gamma\gamma$) decays, with at least one
photon in the barrel. The invariant masses of the photon pairs for $\pi^{0}$ and $\eta$
candidates must be in the ranges $[0.115, 0.150]$~GeV/$c^{2}$ and
$[0.490, 0.580]$~GeV/$c^{2}$, respectively, which are about three times the
resolution of the detector. A kinematic fit that constrains the $\gamma\gamma$
invariant mass to the $\pi^{0}$ or $\eta$ nominal mass~\cite{PDG} is performed
to improve the mass resolution. The $\chi^2$ of the kinematic fit is required
to be less than 30. The $\eta^{\prime}$ candidates are formed from the
$\pi^{+}\pi^{-}\eta$ combinations with an invariant mass within a range of
$[0.946, 0.970]$~GeV/$c^{2}$.

Seven tag modes are used to reconstruct the tag $D_{s}^{-}$ candidate and
its mass~($M_{\rm tag}$) is required to fall within the mass window listed
in Table~\ref{tab:tag-cut}. The recoiling mass of the tag $D_{s}^{-}$ candidate
\begin{eqnarray}
\begin{aligned}
	M_{\rm rec} = \left({\left(\sqrt{s}-\sqrt{|\vec{p}_{D_{s}}|^{2}+m_{D_{s}}^{2}}\right)^{2}-|\vec{p}_{D_{s}}|^{2}}\right)^{1/2}\label{eq:mrec}
\end{aligned}\end{eqnarray} 
is calculated in the $e^+e^-$ center-of-mass system, 
where 
$\vec{p}_{D_{s}}$ is the momentum of the $D_{s}^{-}$ candidate in the 
$e^+e^-$ center-of-mass frame and $m_{D_{s}}$ is the known $D_{s}^{-}$ 
mass~\cite{PDG}. The value of $M_{\rm rec}$ is required to be within the region 
listed in Table~\ref{energe}. 

\begin{table}[htbp]
 \renewcommand\arraystretch{1.25}
 \centering
 \caption{Requirements on $M_{\rm tag}$ for various tag modes.
 }\label{tab:tag-cut}
     \begin{tabular}{lc}
        \hline
        Tag mode                                     & Mass window (GeV/$c^{2}$) \\
        \hline
        $D_{s}^{-} \to K_{S}^{0}K^{-}$               & [1.948, 1.991]            \\
        $D_{s}^{-} \to K^{+}K^{-}\pi^{-}$            & [1.950, 1.986]            \\
        $D_{s}^{-} \to K_{S}^{0}K^{+}\pi^{0}$        & [1.946, 1.987]            \\
        $D_{s}^{-} \to K_{S}^{0}K^{-}\pi^{-}\pi^{+}$ & [1.958, 1.980]            \\
        $D_{s}^{-} \to K_{S}^{0}K^{+}\pi^{-}\pi^{-}$ & [1.953, 1.983]            \\
        $D_{s}^{-} \to \pi^{-}\eta^{\prime}$
                                                     & [1.940, 1.996]            \\
        $D_{s}^{-} \to K^{-}\pi^{+}\pi^{-}$          & [1.953, 1.986]            \\ 
        \hline
      \end{tabular}
\end{table}

\section{Amplitude analysis}
\label{Amplitude-Analysis}
\subsection{Further selections}
\label{AASelection}
The following selection criteria are further applied in order to obtain data
samples with high purity for the amplitude analysis. The selection criteria
discussed in this section are not used in the BF measurement.

An eight-constraint kinematic fit is performed to select photon from $D_{s}^{*\pm}$  
decays and the best DT candidates assuming $D_{s}^{-}$ candidates
decaying to one of the tag modes and $D_{s}^{+}$ decaying to the signal mode
with two hypotheses: the signal $D_s^{+}$ comes from a $D_s^{*+}$ or the tag
$D_s^{-}$ comes from a $D_s^{*-}$. In this kinematic fit, the total four-momentum is constrained to
the initial four-momentum of the $e^+e^-$ system, and the invariant masses of
$(\gamma\gamma)_{\pi^{0}}$, $(\pi^{+}\pi^{-})_{K_S^0}$, tag $D_{s}^{-}$, and
$D_{s}^{*+(-)}$ candidates are constrained to the corresponding known
masses~\cite{PDG}. The best combination is chosen with the minimum $\chi^2$.
After the selection, an additional constraint of the signal $\pi^+ \pi^0 \pi^0$ invariant 
mass to the known $D_s^+$ mass is added and the updated four-momenta of final-state particles from the
kinematic fit are used for the amplitude analysis in order to ensure that all
candidates fall within the phase-space boundary.

The energy of the transition photon from $D_s^{*+}\to \gamma D_s^{+}$ is
required to be smaller than 0.18~GeV. The recoiling mass against this photon
and the signal $D_s^{+}$ is required to fall in the range
$[1.952, 1.995]$~GeV/$c^2$. The $D_s^+\to \pi^+\pi^0\eta$ decay contributes to
the background when $\pi^0\eta$ is misreconstructed as $\pi^0\pi^0$.
This background is reduced via an ``$\eta$'' veto to reject events which
simultaneously satisfy $|M_{\gamma_1 \gamma_3}-M_{\eta}|< 10$~MeV/$c^2$ and
$|M_{\gamma_2 \gamma_4}-M_{\pi^0}|< 20$~MeV/$c^2$, where
$M_{\gamma_1 \gamma_3}$ and $M_{\gamma_2 \gamma_4}$ are the invariant masses of any combinations of the
photons used to reconstruct the two $\pi^0s$ in the signal decay. There is also background originating from $D^0\to K^-\pi^+\pi^0_1$ versus
$\bar{D}^0\to K^+\pi^-\pi^0_2$ decays, where the $\pi^0$ from $D^0$ is denoted as
$\pi^0_1$ and that from $\bar{D}^0$ as $\pi^0_2$.
It fakes $D_s^+ \to \pi^+\pi^0_1\pi^0_2$ versus
$D_s^- \to K^+K^-\pi^-$ ($D_s^- \to \pi^-\pi^0_1\pi^0_2$ versus
$D_s^+ \to K^+K^-\pi^+$) decays by exchanging $K^-$ and $\pi^0_2$ ($K^+$ and
$\pi^0_1$).  This background is excluded by rejecting
events which simultaneously satisfy
$|M_{K^-\pi^+\pi^0_1(\pi^0_2)}-m_{D^{0}}|< 40$~MeV/$c^2$ and
$|M_{K^+\pi^+\pi^0_2(\pi^0_1)}-m_{D^{0}}|< 40$~MeV/$c^2$, where $m_{D^{0}}$ is
the known $D^0$ mass~\cite{PDG}.
A $K^0_S\rightarrow \pi^0\pi^0$ mass veto,
$M_{\pi^0\pi^0}\notin (0.458, 0.520)$~GeV/$c^2$,
is also applied on the signal $D_{s}^+$ to remove the
peaking background $D_s^{+}\to K_{S}^{0}\pi^+$.

\begin{figure*}[!htbp]
  \centering
  \includegraphics[width=0.45\textwidth]{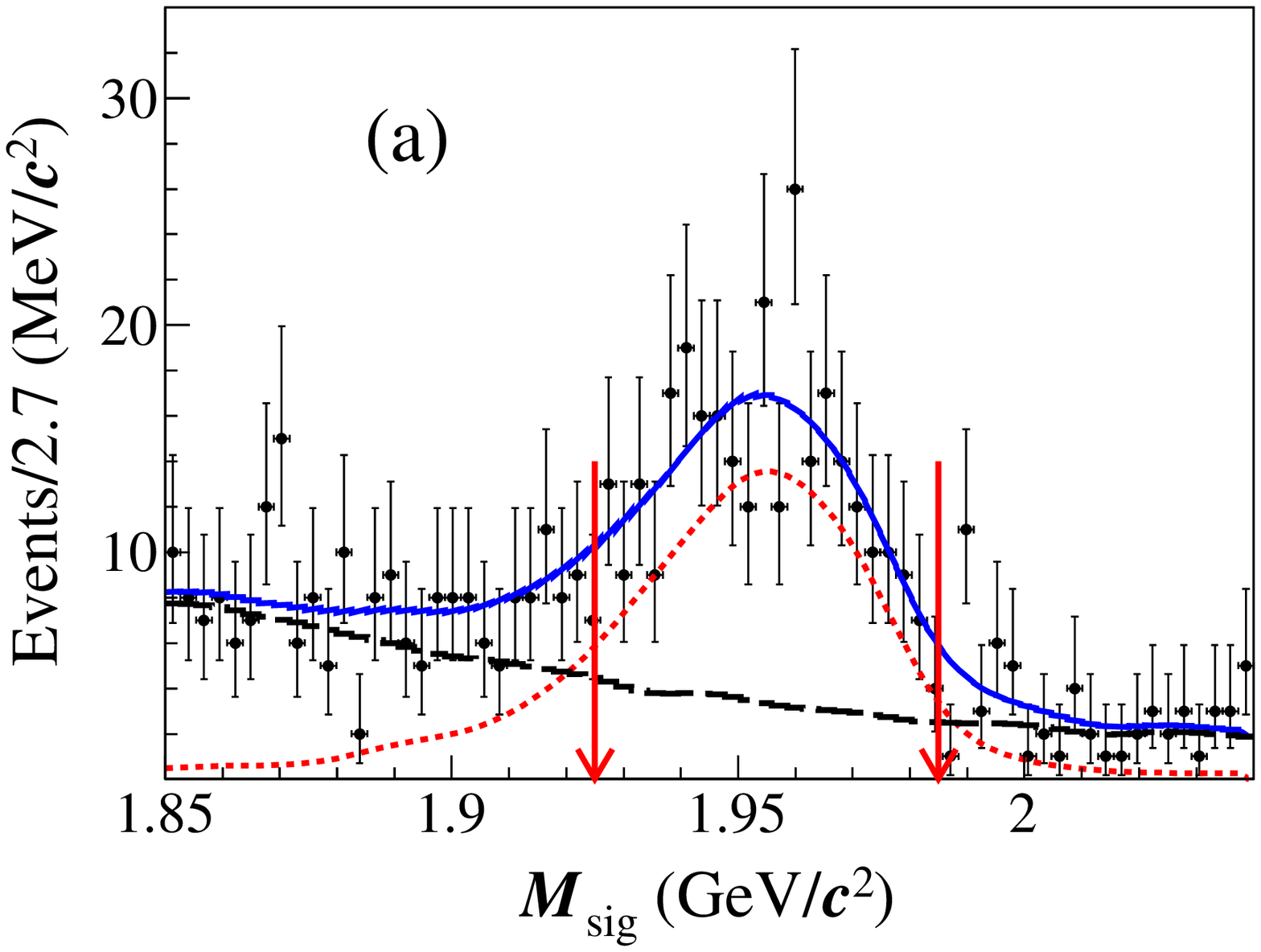}
  \includegraphics[width=0.45\textwidth]{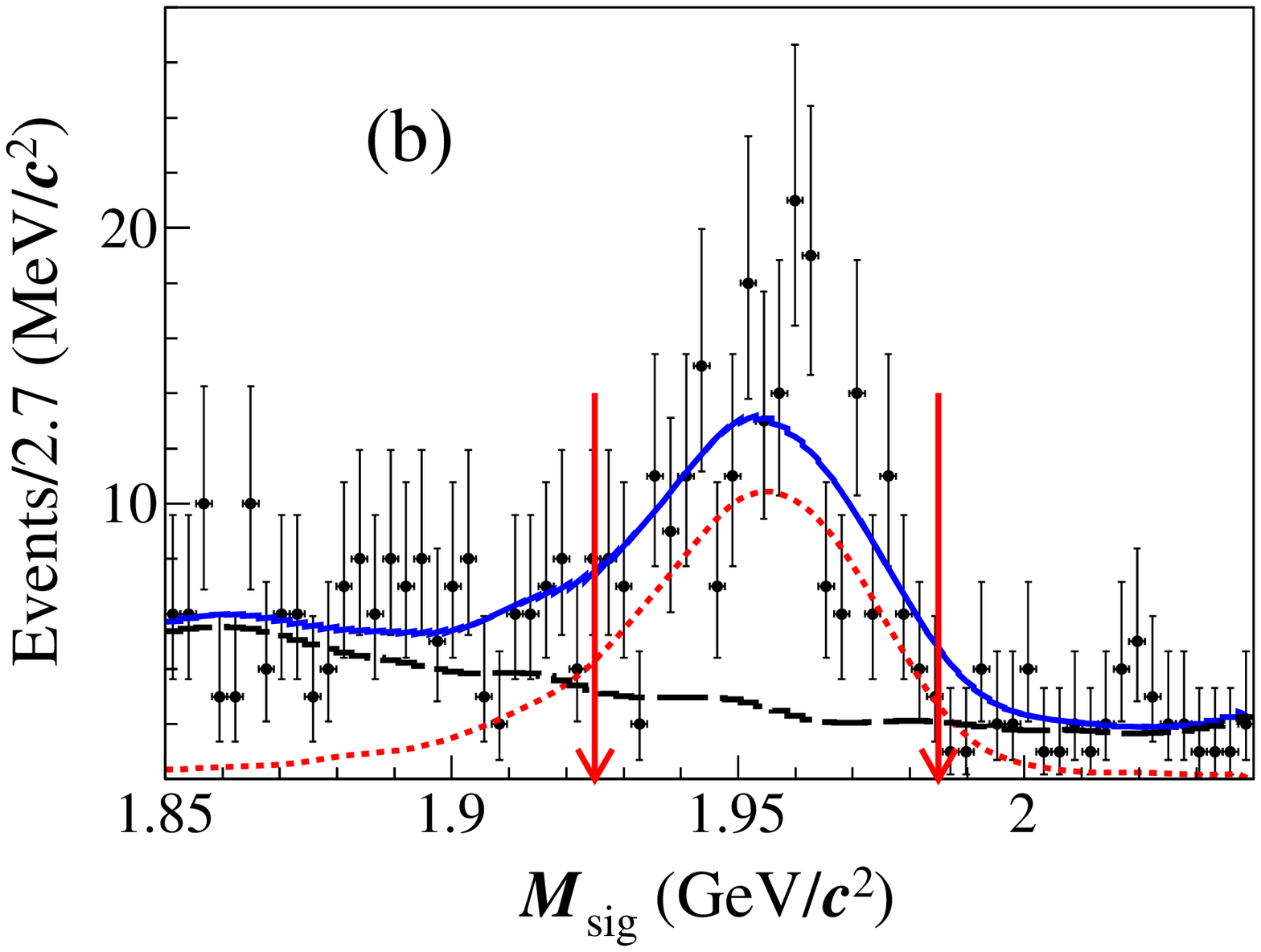}
  \caption{
    Fits to the $M_{\rm sig}$ distributions of the data samples at $\sqrt{s}=$
    (a) 4.178~GeV and (b) 4.189-4.226~GeV. The black points with error bars are
    data. The blue solid lines are the fit results. The red dotted and the black
    dashed lines are the fitted signal and background components, respectively.
    The red arrows indicate the signal regions.
  } \label{fig:fit_Ds}
\end{figure*}

Figure~\ref{fig:fit_Ds} shows the fits to the invariant-mass distributions of the
$D_s^{+}$ candidates reconstructed in the signal mode, $M_{\rm sig}$, for the two data samples. The signal is
described by a MC-simulated shape convolved with a Gaussian resolution
function and the background is described by a simulated shape based on inclusive MC samples.
Finally, a mass window $[1.925, 1.985]$~GeV/$c^2$ is applied. There are 322 and 250 events retained for the amplitude analysis
with purities of $(78.9\pm2.3)\%$ and $(75.6\pm2.9)\%$ for the
data samples at $\sqrt{s}=4.178$~GeV and 4.189-4.226~GeV, respectively.

\subsection{Fit method}
The intermediate-resonant composition is determined by an unbinned
maximum-likelihood fit to data. The likelihood function $\mathcal{L}$ is
constructed with a signal-background combined probability density
function~(PDF), which depends on the momenta of the three final state particles:
\begin{eqnarray}\begin{aligned}
  \mathcal{L} = \prod_{i=1}^{2}\prod_{k=1}^{N_{D, i}}\left[w^{i}f_{S}(p_{j}^{k})+(1-w^{i})f_{B}(p_{j})\right]\,,  \label{likelihood3}
\end{aligned}\end{eqnarray}
where $i$ and $j$ indicate the data sample groups and the final-state particles,
respectively, $N_{D,i}$ is the number of candidate events in the data $i$,
$f_{S}$~($f_{B}$) is the signal~(background) PDF and $w$ is the purity of
signal.

The signal PDF is written as 
\begin{eqnarray}\begin{aligned}
	f_{S}(p_{j}) = \frac{\epsilon(p_{j})\left|\mathcal{A}(p_{j})\right|^{2}R_{3}}{\int \epsilon(p_{j})\left|\mathcal{A}(p_{j})\right|^{2}R_{3}\,dp_{j}}\,, \label{signal-PDF}
\end{aligned}\end{eqnarray}
where $\epsilon(p_{j})$ is the detection efficiency modeled by a RooNDKeysPdf derived from phase space MC sample, 
$\mathcal{A}(p_{j})$ represents the total amplitude, and $R_{3}$ is the standard element of three-body phase
space. The isobar formulism is used to model the total amplitude. The total
amplitude is the coherent sum of individual amplitudes of intermediate
processes, $\mathcal{A}=\sum \rho_{n}e^{i\phi_{n}}\mathcal{A}_{n}$ where magnitude $\rho_{n}$ and
phase $\phi_{n}$ are the free parameters to be determined by data.
The amplitude of the $n^{\rm th}$ intermediate process~($\mathcal{A}_{n}$) is given by
\begin{eqnarray}
\begin{aligned}
	\mathcal{A}_{n} = P_{n}S_{n}F_{n}^{r}F_{n}^{D}\,, \label{base-amplitude}
\end{aligned}\end{eqnarray}
where $S_{n}$ is the spin factor (Sec.~\ref{sec:spinfactor}); $F_{n}^{r}$ and
$F_{n}^{D}$ are the Blatt-Weisskopf barriers of the intermediate state and the
$D_{s}^{+}$ meson, respectively (Sec.~\ref{sec:barrier}); $P_{n}$ is the
propagator of the intermediate resonance (Sec.~\ref{sec:propagator}). The two
identical final state $\pi^0$'s are symmetrized in the model.

The background PDF is given by
\begin{eqnarray}\begin{aligned}
  f_{B}(p_{j}) = \frac{\epsilon(p_{j})B_{\epsilon}(p_{j})R_{3}}{\int \epsilon(p_{j})B_{\epsilon}(p_{j})R_{3}\,dp_{j}}\,,\label{bkg-PDF}
\end{aligned}\end{eqnarray}
where $B_{\epsilon}(p_{j})=B(p_{j})/\epsilon(p_{j})$ is the efficiency-corrected
background shape. The background events in the signal region from the generic
MC sample are used to derive the background shape $B(p_{j})$ with RooNDKeysPdf~\cite{Verkerke}.
RooNDKeysPdf is a kernel estimation method~\cite{Cranmer} implemented in RooFit~\cite{Verkerke}
which models the distribution of an input dataset as a superposition of Gaussian kernels.
The $M_{\pi^+\pi^0}$ and $M_{\pi^0\pi^0}$
distributions of events outside the $M_{\rm sig}$ signal region between the data and the
generic MC samples are compared to check validity of the background from
the generic MC samples. The distributions of background events from the generic MC samples
within and outside the $M_{\rm sig}$ signal region are also examined.
They are found to be compatible within statistical uncertainties.
Note that the $\epsilon(p_{j})$ term in Eq.~(\ref{bkg-PDF}) is explicitly written out as it is independent of the fitted variables and is dropped during the log-likelihood fit.
The normalization integral terms in the signal and background PDF are handled
by MC integration,
\begin{eqnarray}\begin{aligned}
  \int \epsilon(p_{j}) X(p_{j}) R_{3}\,dp_{j} \approx
\frac{1}{N_{\rm G}}\sum_{k}^{N_{\rm M}} \frac{ X(p_{j}^{k}) }{\left|\mathcal{M}^{g}(p_{j}^{k})\right|^{2}}\,, \label{MC-intergral}
\end{aligned}\end{eqnarray}
where $X(p_{j})$ is $|\mathcal{A}(p_{j})|^2$ or $B_{\epsilon}(p_{j})$, $k$ is the index of
the $k^{\rm th}$ event, $N_{\rm G}$ is the number of the generated MC events and $N_{\rm M}$ is the number of the selected MC events.
The $D_s^+$ meson in the MC samples used here decays to $\pi^+\pi^0\pi^0$
according to the PDF $\mathcal{M}^{g}(p_{j})$, while the $D_s^-$ meson decays into one of
the tag modes. These MC samples are generated with different $\sqrt{s}$
according to the luminosities and cross sections, and satisfy all selection
criteria as those of the data samples. At the beginning, a
preliminary PDF is used, and then a recursive process is performed until the
result converges. To account for any bias caused by differences in PID or
tracking efficiencies between data and MC simulation, each signal MC event is
weighted with a ratio, $\gamma_{\epsilon}(p)$, of the efficiency of data to that
of MC simulation and the MC integration then becomes
\begin{eqnarray}\begin{aligned}
    &\int \epsilon(p_{j}) X(p_{j}) R_{3}\,dp_{j} \approx
&\frac{1}{N_{\rm G}} \sum_{k}^{N_{\rm M}} \frac{ X(p_{j}^{k}) \gamma_{\epsilon}(p_{j}^{k})}{\left|\mathcal{M}^{g}(p_{j}^{k})\right|^{2}}\,.
\label{MC-intergral-corrected}
\end{aligned}\end{eqnarray}

\subsubsection{Spin factors}\label{sec:spinfactor}
The spin-projection operators are defined as~\cite{covariant-tensors}
\begin{eqnarray}
\begin{aligned}
  P^{(1)}_{\mu\mu^{\prime}}(a) &= -g_{\mu\mu^{\prime}}+\frac{p_{a,\mu}p_{a,\mu^{\prime}}}{p_{a}^{2}}\,,\\
  P^{(2)}_{\mu\nu\mu^{\prime}\nu^{\prime}}(a) &= \frac{1}{2}(P^{(1)}_{\mu\mu^{\prime}}(a)P^{(1)}_{\nu\nu^{\prime}}(a)+P^{(1)}_{\mu\nu^{\prime}}(a)P^{(1)}_{\nu\mu^{\prime}}(a))\\
  &-\frac{1}{3}P^{(1)}_{\mu\nu}(a)P^{(1)}_{\mu^{\prime}\nu^{\prime}}(a)\,.
 \label{spin-projection-operators}
\end{aligned}
\end{eqnarray}
The quantities $p_a$, $p_b$, and $p_c$ are the momenta of particles $a$,
$b$, and $c$, respectively, and $r_a = p_b-p_c$.
The covariant tensors are given by
\begin{eqnarray}
\begin{aligned}
    \tilde{t}^{(1)}_{\mu}(a) &= -P^{(1)}_{\mu\mu^{\prime}}(a)r^{\mu^{\prime}}_{a}\,,\\
    \tilde{t}^{(2)}_{\mu\nu}(a) &= P^{(2)}_{\mu\nu\mu^{\prime}\nu^{\prime}}(a)r^{\mu{\prime}}_{a}r^{\nu^{\prime}}_{a}\,.\\
\label{covariant-tensors}
\end{aligned}
\end{eqnarray}
The spin factors for $S$, $P$, and $D$ wave decays are
\begin{eqnarray}
\begin{aligned}
    S &= 1\,, &(S\ \text{wave}), &\\
    S &= \tilde{T}^{(1)\mu}(D_{s}^{\pm})\tilde{t}^{(1)}_{\mu}(a)\,,         &(P\ \text{wave}),\\
    S &= \tilde{T}^{(2)\mu\nu}(D_{s}^{\pm})\tilde{t}^{(2)}_{\mu\nu}(a)\,,         &(D\ \text{wave}),
\label{spin-factor}
\end{aligned}
\end{eqnarray}
where the $\tilde{T}^{(l)}$ factors have the same definition as $\tilde{t}^{(l)}$. The
tensor describing the $D_{s}^{+}$ decay is denoted by $\tilde{T}$ and
that of the $a$ decay is denoted by $\tilde{t}$.

\subsubsection{Blatt-Weisskopf barrier factors}\label{sec:barrier}
For the process $a \to bc$, the Blatt-Weisskopf barrier $F_L(p_j)$~\cite{PhysRevD.83.052001} is
parameterized as a function of the angular momentum $L$ and the momentum $q$ of
the final-state particle $b$ or $c$ in the rest system of $a$,
\begin{eqnarray}
\begin{aligned}
 F_{L=0}(q)&=1,\\
 F_{L=1}(q)&=\sqrt{\frac{z_0^2+1}{z^2+1}},\\
 F_{L=2}(q)&=\sqrt{\frac{z_0^4+3z_0^2+9}{z^4+3z^2+9}}\,,
\end{aligned}
\end{eqnarray}
where $z=qR$ and $z_0=q_0R$. The effective radius of the barrier $R$ is fixed
to 3.0~GeV$^{-1}$ for the intermediate resonances and 5.0~GeV$^{-1}$ for the
$D_s^+$ meson.

\subsubsection{Propagator}\label{sec:propagator}
The intermediate resonances $f_2(1270)$ and $f_0(1370)$ are
parameterized as relativistic Breit-Wigner functions,
\begin{eqnarray}\begin{aligned}
	P = \frac{1}{(m_{0}^{2} - s_{a} ) - im_{0}\Gamma(s_{a})}\,,\; 
	\Gamma(s_{a}) = \Gamma_{0}\left(\frac{q}{q_{0}}\right)^{2L+1}\left(\frac{m_{0}}{s_{a}}\right)\left(\frac{F_{L}(q)}{F_{L}(q_{0})}\right)^{2}\,, 
  \label{RBW}
\end{aligned}\end{eqnarray}
where $s_{a}$ denotes the invariant-mass squared of the two final-state particles considered;
$m_{0}$ and $\Gamma_{0}$ are the mass and the width of the intermediate
resonance, respectively, and are fixed to the PDG values~\cite{PDG}.

The $f_{0}(980)$ resonance is represented by the Flatt$\acute{\rm e}$ formula~\cite{flatte_f0},
\begin{equation}
	P_{f_0(980)}= \frac{1}{m_{f_{0}(980)}^{2} - s_{\pi^{0}\pi^{0}} - i(g_{1}\rho_{\pi\pi}(s_{\pi^{0}\pi^{0}}) + g_{2}\rho_{K\bar{K}}(s_{\pi^{0}\pi^{0}}))}, \label{Flatte}
\end{equation}
where $s_{\pi^{0}\pi^{0}}$ is the $\pi^0\pi^0$ invariant-mass squared and
$g_{1,2}$ are coupling constants to the corresponding final states. The
parameters are fixed to $g_{1}=0.165$~GeV$/c^2$, $g_{2}/g_{1}=4.21$
and $m_{f_{0}(980)}=955$~MeV/$c^{2}$, as reported in Ref.~\cite{flatte_f0}.
The Lorentz invariant phase-space factors $\rho_{\pi\pi}(s)$ and $\rho_{K\bar{K}}(s)$
are given by
\begin{eqnarray}
\begin{aligned}
    \rho_{\pi\pi}&=\frac{2}{3}\sqrt{1-\frac{4m^{2}_{\pi^{\pm}}}{s}}+\frac{1}{3}\sqrt{1-\frac{4m^{2}_{\pi^{0}}}{s}}\,,\\
    \rho_{K\bar{K}}&=\frac{1}{2}\sqrt{1-\frac{4m^{2}_{K^{\pm}}}{s}}+\frac{1}{2}\sqrt{1-\frac{4m^{2}_{K^{0}}}{s}}\,,
\end{aligned}
\end{eqnarray}
where $m_{\pi^{\pm}}$, $m_{\pi^{0}}$, $m_{K^{\pm}}$, and $m_{K^{0}}$ are the known masses of
$\pi^{\pm}$, $\pi^{0}$, $K^{\pm}$, and $K^{0}$, respectively~\cite{PDG}.
The $f_{0}(500)$ resonance is also an amplitude candidate, and is described by a
relativistic Breit-Wigner function or the Bugg lineshape~\cite{ref:bugg}.

\subsection{Fit results}
\begin{figure}[htbp]
    \centering
     \includegraphics[width=0.45\textwidth]{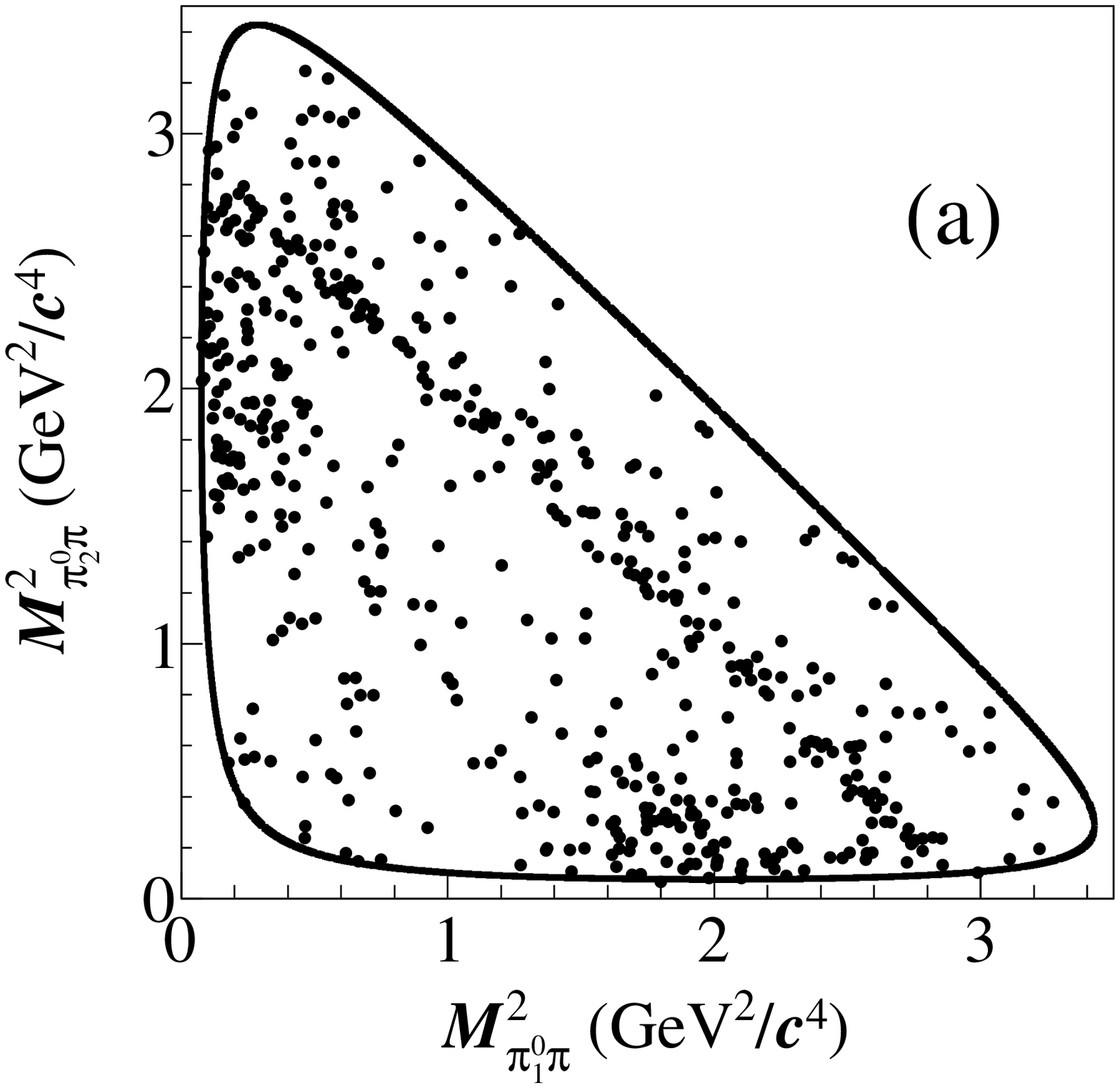}
     \includegraphics[width=0.45\textwidth]{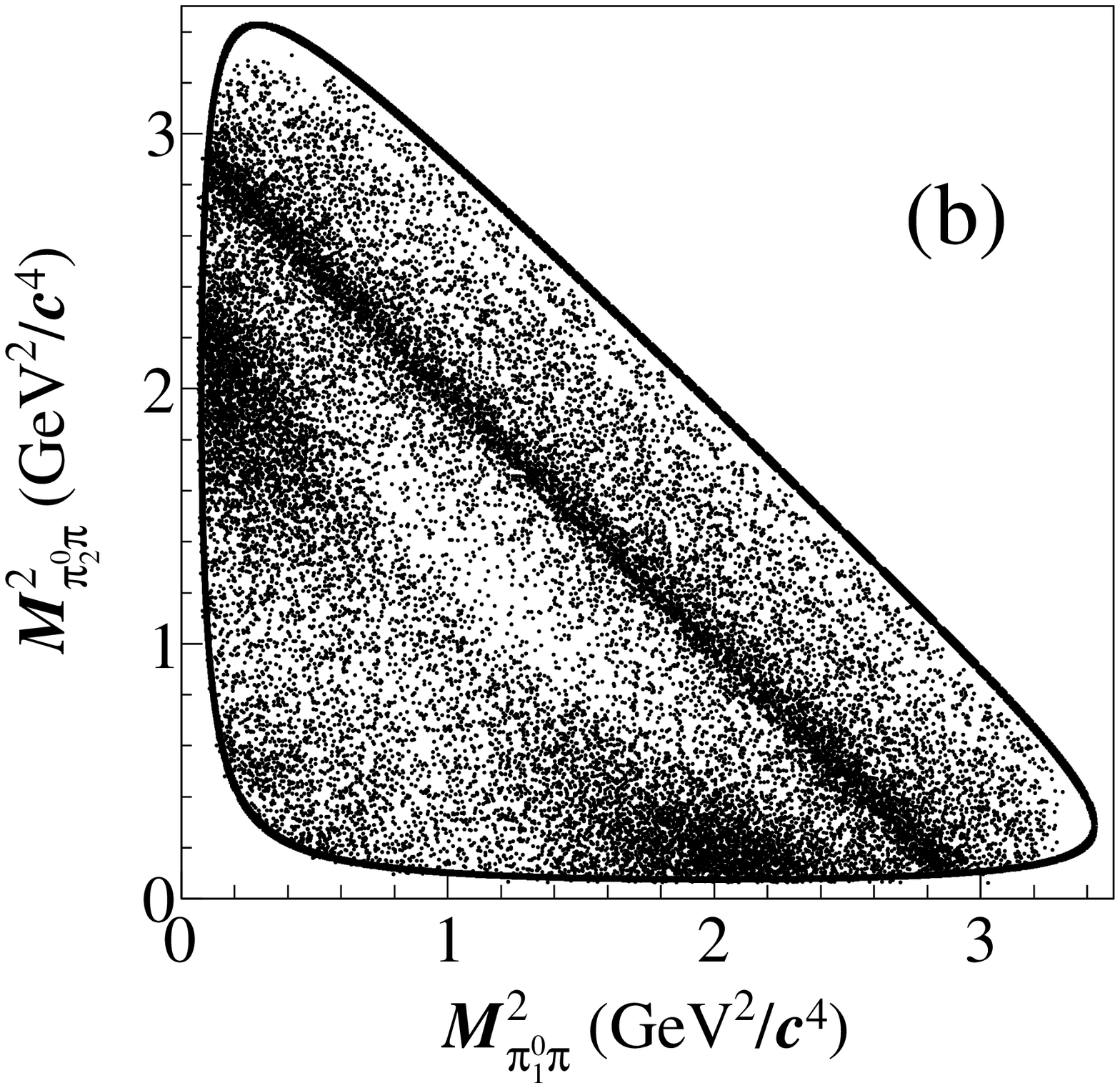}
  \caption{The Dalitz plot of $M^{2}_{\pi^+\pi^0}$ versus~$M^{2}_{\pi^+\pi^0}$
    for (a) the data sample and (b) the signal MC sample generated based on
    the results of the amplitude analysis at $\sqrt{s}= 4.178$-$4.226$~GeV,
    symmetrized for the indistinguishable $\pi^{0}$'s.}
  \label{dalitz}
\end{figure}

The Dalitz plot of $M^{2}_{\pi^+\pi^0}$ versus~$M^{2}_{\pi^+\pi^0}$ for the
data samples is shown in Fig.~\ref{dalitz}(a) and that for the signal MC samples
generated based on the results of the amplitude analysis is shown in
Fig.~\ref{dalitz}(b).
In the fit, the magnitude and phase of the reference amplitude
$D_{s}^{+} \to f_0(980)\pi^+$ are fixed to 1.0 and 0.0, respectively, while
those of other amplitudes are left floating. The masses and widths of all resonances
are fixed to the corresponding PDG averages~\cite{PDG}, and $w^{i}$
are fixed to the purities discussed in Sec.~\ref{AASelection}. The systematic
uncertainties associated with these fixed parameters are considered by
repeating the fit after variation of the fixed parameters according to their uncertainties.

Besides the dominant amplitudes $D_{s}^{+} \to f_0(980)\pi^{+}$,
$D_{s}^{+} \to f_0(1370)\pi^{+}$, and $D_{s}^{+} \to f_2(1270)\pi^{+}$,
we have tested all possible intermediate resonances including $\rho(1450)$, $f_0(1500)$,
$\rho(1700)$, $(\pi\pi)_{S}$, $(\pi\pi)_{P}$, $(\pi\pi)_{D}$ etc.,
where the subscript denotes a relative $S$ ($P$ or $D$) wave between final-state particles. We have also examined all possible combinations of these
intermediate resonances to check their significances, correlations, and
interferences.
By requiring a significance larger than $3\sigma$,
eventually,  $D_{s}^{+} \to f_0(980)\pi^{+}$,
$D_{s}^{+} \to f_0(1370)\pi^{+}$, $D_{s}^{+} \to f_2(1270)\pi^{+}$,
$D_{s}^{+} \to \pi^{+}(\pi^0\pi^0)_{D}$, and
$D_{s}^{+} \to (\pi^{+}\pi^0)_{D}\pi^0$ are chosen for the nominal set.
Note that $D_{s}^{+} \to f_0(500)\pi^{+}$ is tested but it has a significance
less than $2\sigma$.

In the calculation of fit fractions (FFs) for individual amplitudes, the
phase-space MC truth information is involved with neither detector acceptance
nor resolution. The FF for the $n^{\rm th}$ amplitude is defined as
\begin{eqnarray}\begin{aligned}
  {\rm FF}_{n} = \frac{\sum^{N_{\rm gen}} \left|c_{n}\mathcal{A}_{n}\right|^{2}}{\sum^{N_{\rm gen}} \left|\mathcal{A}\right|^{2}}\,, \label{Fit-Fraction-Definition}
\end{aligned}\end{eqnarray}
where $N_{\rm gen}$ is the number of the phase-space MC events at generator
level. Interference between the $n^{\rm th}$ and the
$n^{\prime{\rm th}}$ amplitudes (IN) is defined as (for $n<n^{\prime}$ only)
\begin{eqnarray}\begin{aligned}
  {\rm IN}_{nn^{\prime}} = \frac{\sum^{N_{\rm gen}} 2\textrm{Re}[c_{n}c^{*}_{n^{\prime}}\mathcal{A}_{n}\mathcal{A}^{*}_{n^{\prime}}]}{\sum^{N_{\rm gen}} \left|\mathcal{A}\right|^{2}}\,. \label{interferenceFF-Definition}
\end{aligned}\end{eqnarray}
The statistical fluctuations of FFs are obtained by randomly sampling the fit variables according to their fitted values and covariance matrix. The distribution of each FF is fitted with a Gaussian function and the width of
the Gaussian function is defined as the statistical uncertainty of the FF.

\begin{figure*}[!htbp]
 \centering
 \includegraphics[width=0.45\textwidth]{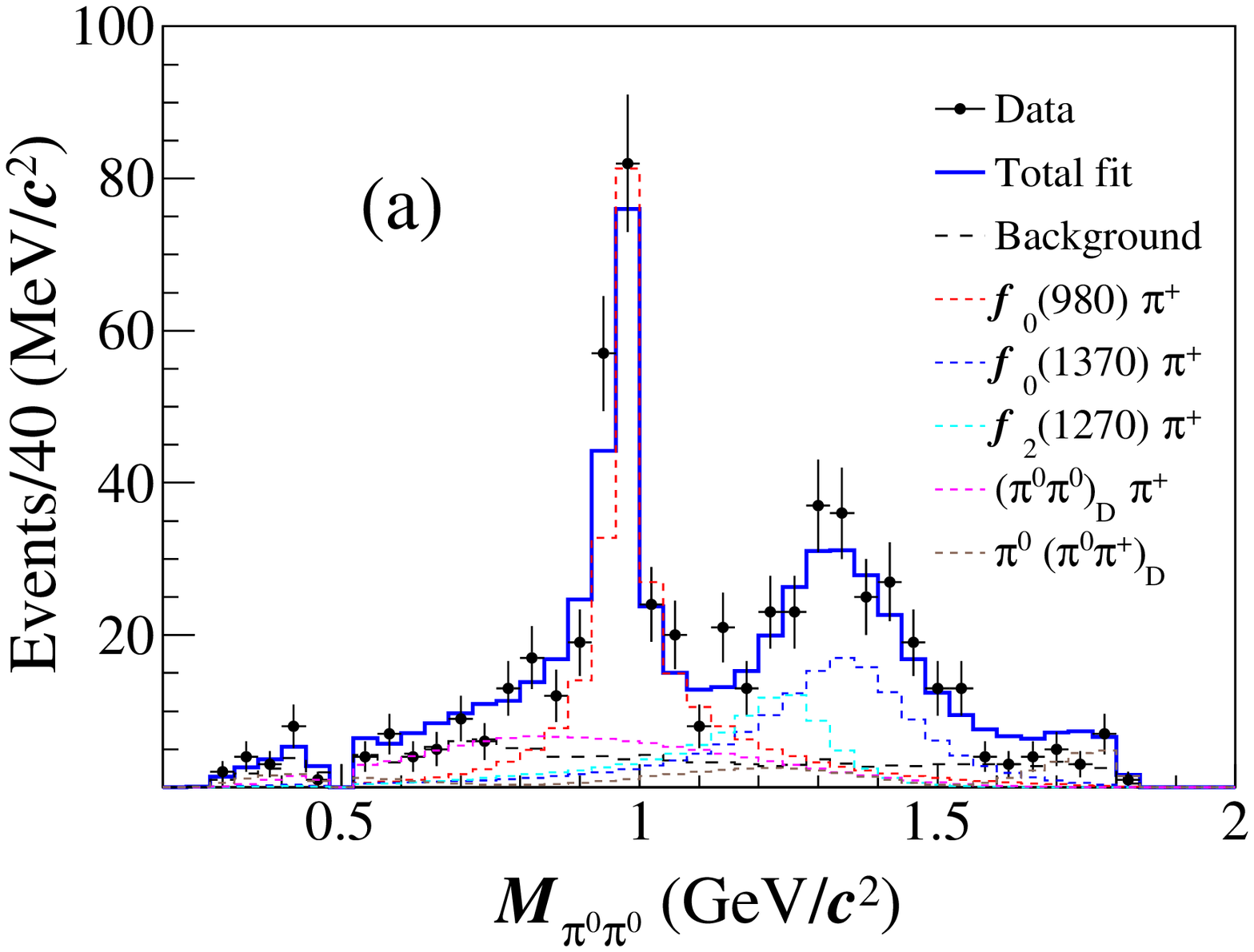}
 \includegraphics[width=0.45\textwidth]{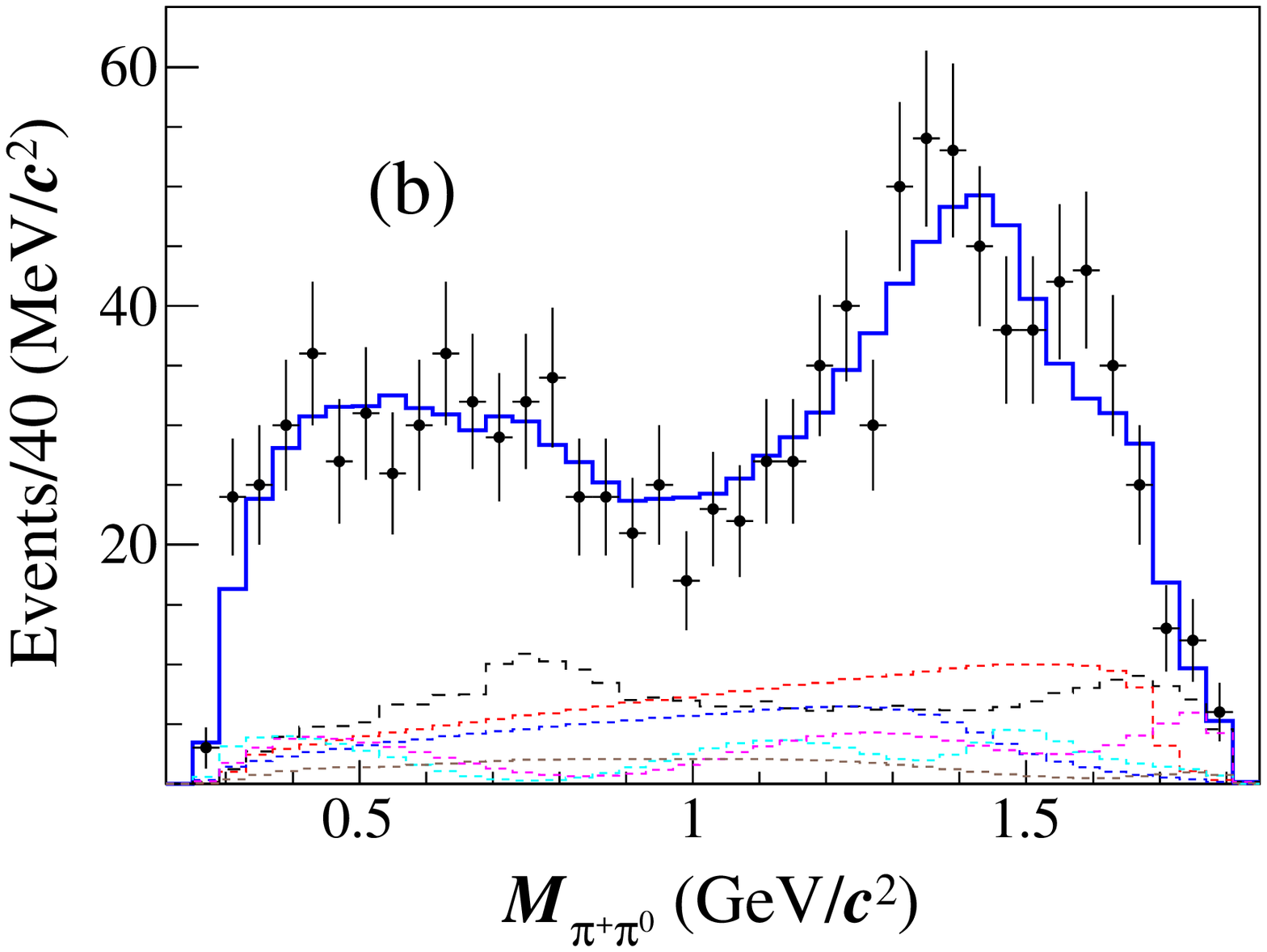}
 \caption{
   The projections of (a) $M_{\pi^0\pi^0}$ and (b) $M_{\pi^+\pi^0}$ from the
   nominal fit. Two $M_{\pi^+\pi^0}$ are calculated and added due to the
   indistinguishable $\pi^0$'s. The data samples 
   are represented by points with error bars, the fit results by the solid blue
   lines, and the background estimated from inclusive MC samples by the black
   dashed lines. Colored dashed lines show the components of the fit model.
   Due to interference effects, the total is not necessarily equal to the sum
   of the components.
}
 \label{dalitz-projection}
\end{figure*}

The phases, FFs, and statistical significances for the amplitudes are listed in
Table~\ref{fit-result}. The interferences between amplitudes are listed in
Table~\ref{fit-interference}. The Dalitz plot projections are shown in
Fig.~\ref{dalitz-projection}. The sum of the FFs is not unity due to interferences between amplitudes.
Other tested amplitudes, but not included in the nominal fit, and
their significances are listed in Table~\ref{tested}.

\begin{table*}[htbp]
  \caption{The phases, FFs, and statistical significances for the amplitudes.
      The first and
      second uncertainties in the phases and FFs are statistical and systematic,
      respectively. The total FF is 111.4$\%$.}
    \label{fit-result}
    \begin{center}
    \begin{tabular}{lccc}
      \hline
      Amplitude                                   & Phase $\phi_n$ (rad)         & FF~(\%)                           &Significance~($\sigma$)\\
      \hline
      $D_{s}^{+} \to f_0(980)\pi^{+}$             & 0.0(fixed)               & $42.0 \pm 4.9 \pm 6.6$            &$>$10 \\
      $D_{s}^{+} \to f_0(1370)\pi^{+}$            & $ -0.7 \pm 0.2 \pm 0.3$  & $25.8 \pm 4.4 \pm 4.8$            &$>$10\\
      $D_{s}^{+} \to f_2(1270)\pi^{+}$            & $ -0.9 \pm 0.3 \pm 0.3$  & $15.0 \pm 4.6 \pm 5.2$            &5.0\\
      $D_{s}^{+} \to \pi^{+}(\pi^0\pi^0)_{D}$ & $ -4.4 \pm 0.2 \pm 0.3$  & $19.1 \pm 5.2 \pm 5.2$            &6.3\\
      $D_{s}^{+} \to (\pi^{+}\pi^0)_{D}\pi^0$ & $ -2.3 \pm 0.2 \pm 0.5$  & $\phantom{0}9.5 \pm 3.4 \pm 5.1$  &3.4\\
      \hline
    \end{tabular}
    \end{center}
\end{table*}
\begin{table*}[htbp]
    \caption{Interference fraction (\%) between amplitudes where 
       the uncertainties are statistical only.}
    \label{fit-interference}
    \begin{center}
    \begin{tabular}{l|cccc}
      \hline
                                    & $f_0(1370)\pi^{+}$ & $f_2(1270)\pi^{+}$       & $\pi^{+}(\pi^0\pi^0)_{D}$ & $(\pi^{+}\pi^0)_{D}\pi^0$\\
      \hline
      $f_0(980)\pi^{+}$             & $-2.6\pm 7.3$       & $\phantom{0}0.0 \pm 0.1$ & $\phantom{0}-0.1 \pm 0.1$     & $\phantom{0}4.1 \pm 2.4$\\
      $f_0(1370)\pi^{+}$            &                    & $-0.1 \pm 0.1$           & $\phantom{00}0.0 \pm 0.1$     & $\phantom{0}9.2 \pm 1.8$\\
      $f_2(1270)\pi^{+}$            &                    &                          & $-15.7 \pm 4.4$               & $-3.3 \pm 1.7$\\
      $\pi^{+}(\pi^0\pi^0)_{D}$ &                    &                          &                               & $-4.6 \pm 2.8$\\
      \hline
    \end{tabular}
    \end{center}
\end{table*}

\begin{table*}[htbp]
  \caption{Significances of amplitudes tested, but not included in the nominal fit.}
    \label{tested}
    \begin{center}
    \begin{tabular}{lc}
      \hline
      Amplitude                                   & Significance~($\sigma$)\\
      \hline
      $D_{s}^{+} \to f_0(500)\pi^{+}$             & 1.5\\
      $D_{s}^{+} \to f_0(1500)\pi^{+}$            & 2.1\\
      $D_{s}^{+} \to \rho(1450)^+\pi^{0}$         & 2.4\\
      $D_{s}^{+} \to \rho^+\pi^{0}$               & 2.0\\
      $D_{s}^{+} \to (\pi^+\pi^0)_{P}\pi^0$   & 1.5\\
      $D_{s}^{+} \to \pi^{+}(\pi^0\pi^0)_{S}$ & 1.3\\
      \hline
    \end{tabular}
    \end{center}
\end{table*}

\subsection{Systematic uncertainties for the amplitude analysis}
\label{sec:PWA-Sys}
The systematic uncertainties for the amplitude analysis are summarized
in Table~\ref{systematic-uncertainties}, with their definitions described below:
\begin{itemize}
\item[\lowercase\expandafter{\romannumeral1}]
  Resonant parameters. The masses and the widths of $f_0(1370)$ and $f_2(1270)$
  are varied by their corresponding uncertainties~\cite{PDG}. The mass and
  coupling constants of the $f_0(980)$ Flatt$\acute{\rm e}$ formula are varied according to
  Ref.~\cite{flatte_f0}. The changes of the phases $\phi$ and FFs are assigned
  as the associated systematic uncertainties.

\item[\lowercase\expandafter{\romannumeral2}]
  $R$ values. 
  The associated systematic uncertainties are estimated by repeating the fit
  procedure by varying the radii of the intermediate state and $D_s^{+}$ mesons
  within 1~GeV$^{-1}$. 
  
\item[\lowercase\expandafter{\romannumeral3}]
  Background estimation.
  First, the purities of signals for the two sample groups, i.e.~$w$ in
  Eq.~(\ref{likelihood3}) are varied by their corresponding statistical
  uncertainties to study uncertainties associated with backgrounds. The
  differences caused by the variation are assigned as the uncertainties.
  Second, an alternative MC-simulated shape is used to examine the uncertainty
  arising from the background shape modeling. Alternative background shapes are extracted with
  the relative fractions of the dominant backgrounds from $e^+e^-\to q\bar{q}$
  and non-$D_{s}^{*\pm}D_{s}^{\mp}$ open-charm processes varied by the
  statistical uncertainties of their cross sections.


\item[\lowercase\expandafter{\romannumeral4}]
  Resonances with significances less than $3\sigma$.
  The corresponding uncertainties are taken to be the differences of the
  phases $\phi$ and FFs with and without the intermediate resonances with
  statistical significances less than $3\sigma$.
  
\item[\lowercase\expandafter{\romannumeral5}]
  Experimental effects.
  To estimate the systematic uncertainty related to the difference
  between MC simulation and data associated with the PID and tracking efficiencies,
  $\gamma_{\epsilon}$ in Eq.~(\ref{MC-intergral-corrected}), the amplitude
  fit is performed varying the PID and tracking efficiencies according to their
  uncertainties. The differences from the nominal results are so tiny that
  this source of systematic uncertainty is negligible.

\end{itemize}

\begin{table*}[tp]
  \renewcommand\arraystretch{1.25}
  \centering
  \caption{Systematic uncertainties on the phase $\phi$ and FF for each
    amplitude in unit of the corresponding statistical uncertainty.  The sources are:
    (\lowercase\expandafter{\romannumeral1}) fixed parameters in the amplitudes,
    (\lowercase\expandafter{\romannumeral2}) the $R$ values,
    (\lowercase\expandafter{\romannumeral3}) background,
    (\lowercase\expandafter{\romannumeral4}) resonances with significances less than $3\sigma$.
}
  \label{systematic-uncertainties}
  \begin{tabular}{lcccccc}
    \hline
    \multirow{2}{*}{Amplitude}&\multicolumn{6}{c}{Source}\cr
    & & \lowercase\expandafter{\romannumeral1} &\lowercase\expandafter{\romannumeral2} &\lowercase\expandafter{\romannumeral3} &\lowercase\expandafter{\romannumeral4} & Total   \\
    \hline
    $D_{s}^{+} \to f_0(980)\pi^{+}$                               & FF     & 1.07 & 0.29 & 0.31 & 0.70 & 1.35 \\
    \hline
    \multirow{2}{*}{$D_{s}^{+} \to f_0(1370)\pi^{+}$}             & $\phi$ & 1.32 & 0.30 & 0.34 & 0.42 & 1.46 \\
                                                                  & FF     & 1.06 & 0.20 & 0.09 & 0.08 & 1.09 \\
    \hline
    \multirow{2}{*}{$D_{s}^{+} \to f_2(1270)\pi^{+}$}             & $\phi$ & 0.56 & 0.09 & 0.23 & 0.85 & 1.05 \\
                                                                  & FF     & 0.93 & 0.53 & 0.36 & 0.16 & 1.14 \\
    \hline
    \multirow{2}{*}{$D_{s}^{+} \to \pi^{+}(\pi^0\pi^0)_{D}$}  & $\phi$ & 0.56 & 0.42 & 0.24 & 1.53 & 1.70 \\
                                                                  & FF     & 0.88 & 0.46 & 0.10 & 0.11 & 1.00 \\
    \hline
    \multirow{2}{*}{$D_{s}^{+} \to (\pi^{+}\pi^0)_{D}\pi^0$}  & $\phi$ & 1.36 & 0.09 & 0.17 & 2.15 & 2.55 \\
                                                                  & FF     & 0.72 & 0.14 & 0.20 & 1.30 & 1.50 \\
    \hline
  \end{tabular}
\end{table*}

\section{Branching fraction measurement}
\label{BFSelection}
In addition to the selection criteria for final-state particles described in
Sec.~\ref{ST-selection}, it is required that $\pi^+$ must have momentum
greater than $100$~MeV/$c$ to remove soft $\pi^+$ from $D^{*+}$ decays. The
best tag candidate with $M_{\rm rec}$ closest to the $D_{s}^{*+}$ known
mass~\cite{PDG} is chosen if there are multiple ST candidates. The data sets
are organized into three sample groups, 4.178~GeV, 4.189-4.219~GeV, and
4.226~GeV, that were acquired during the same year under consistent running
conditions.

\begin{figure*}[htp]
\begin{center}
  \includegraphics[width=0.4\textwidth]{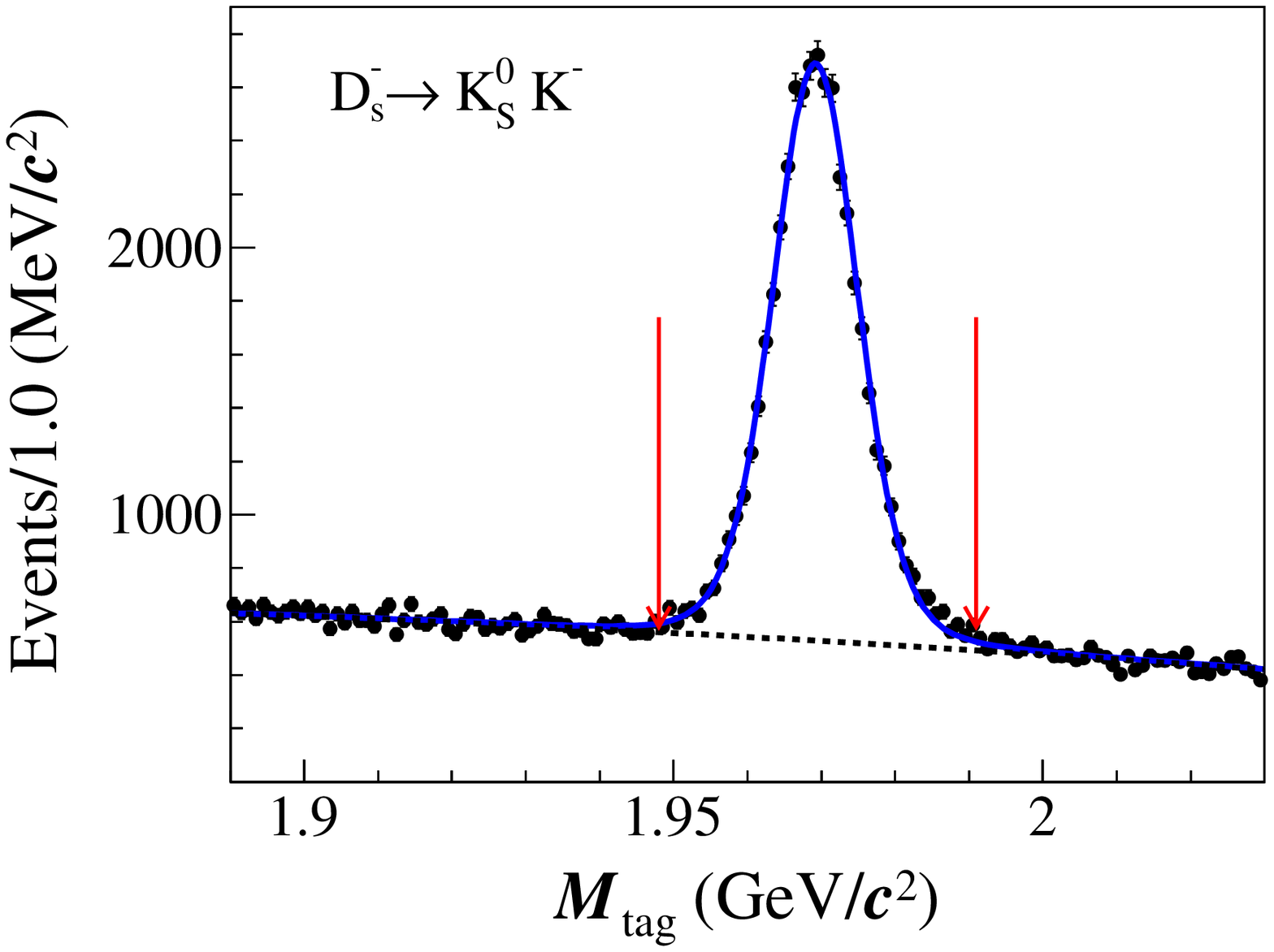}
  \includegraphics[width=0.4\textwidth]{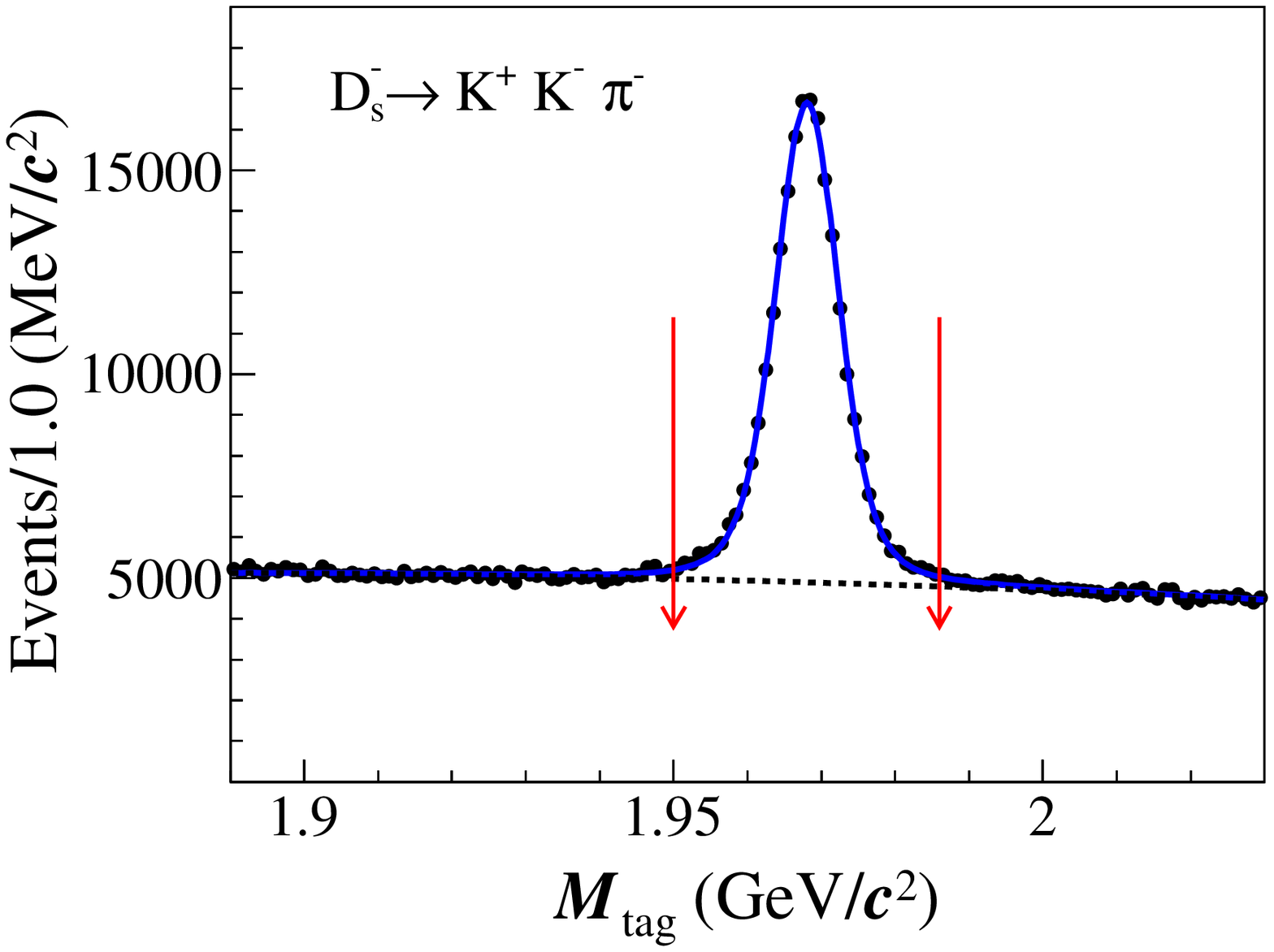}\\
  \includegraphics[width=0.4\textwidth]{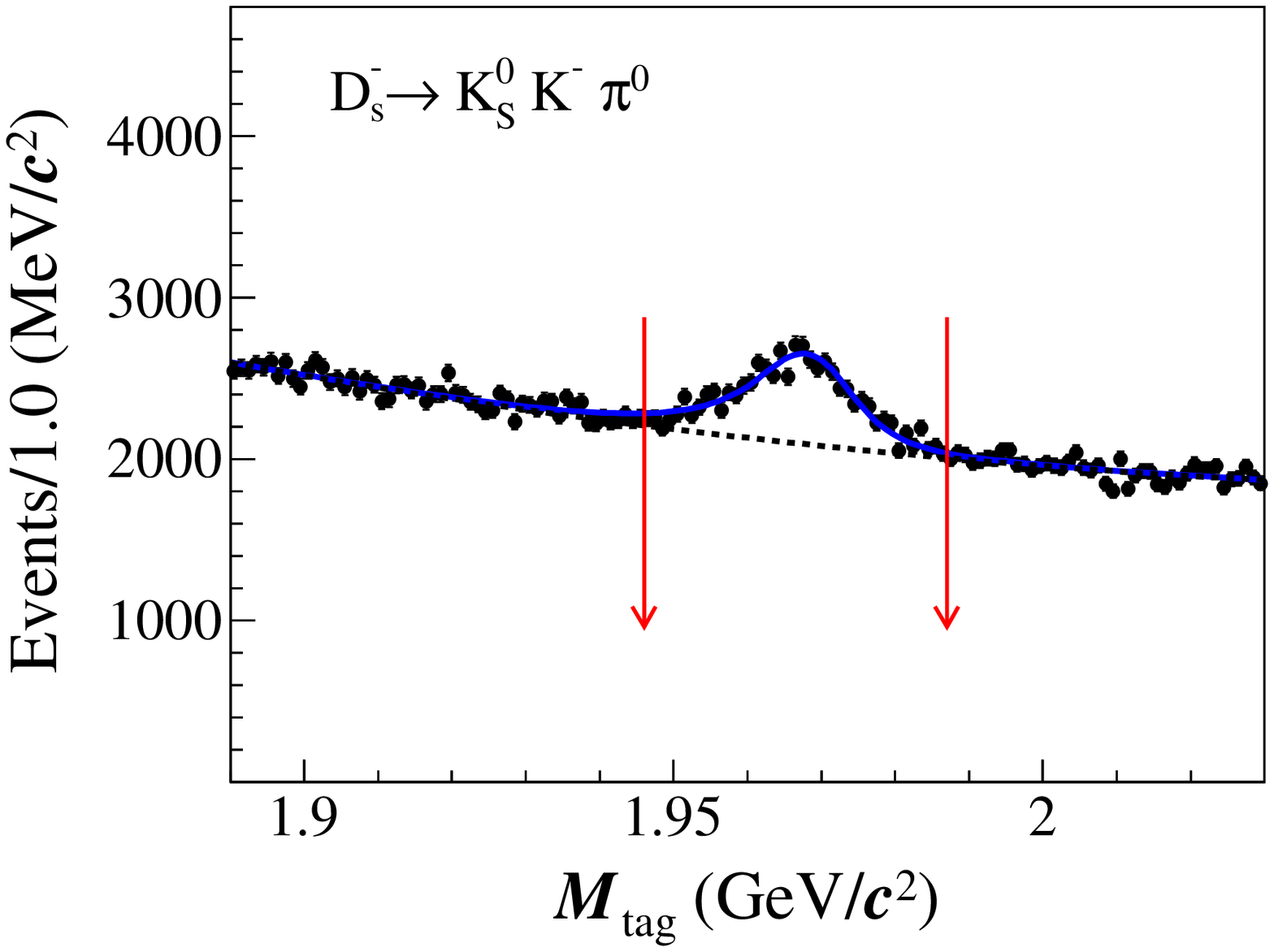}
  \includegraphics[width=0.4\textwidth]{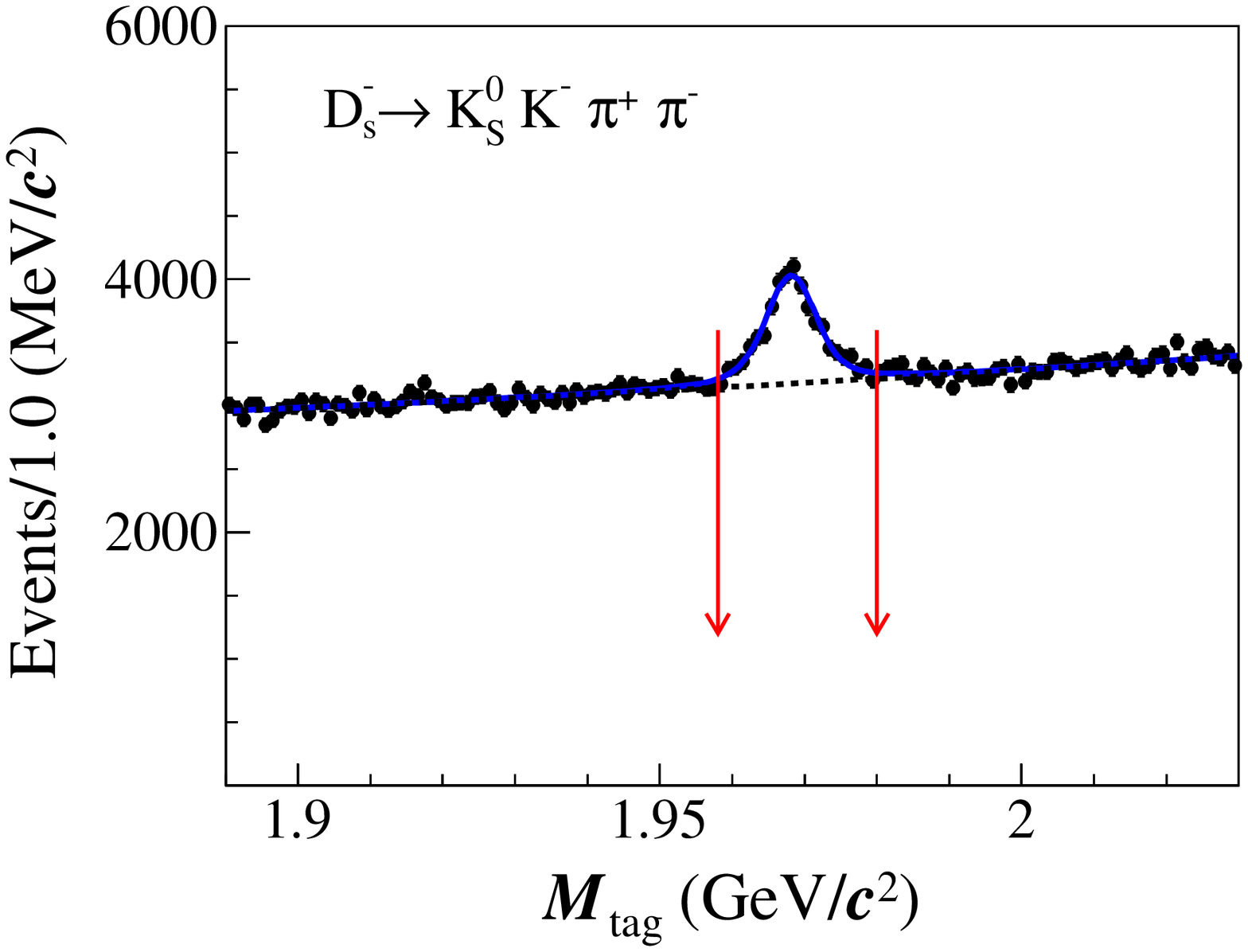}\\
  \includegraphics[width=0.4\textwidth]{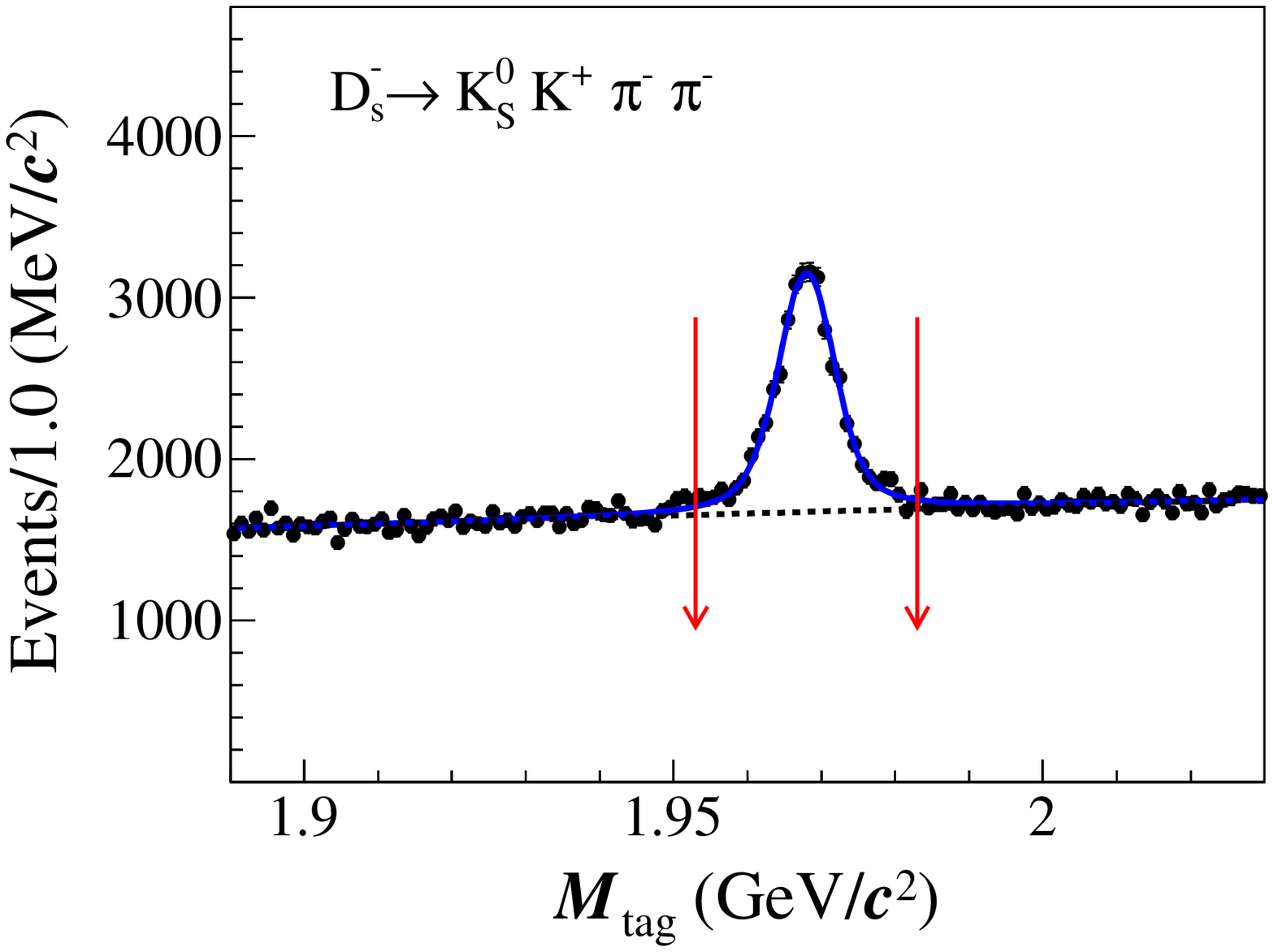}
  \includegraphics[width=0.4\textwidth]{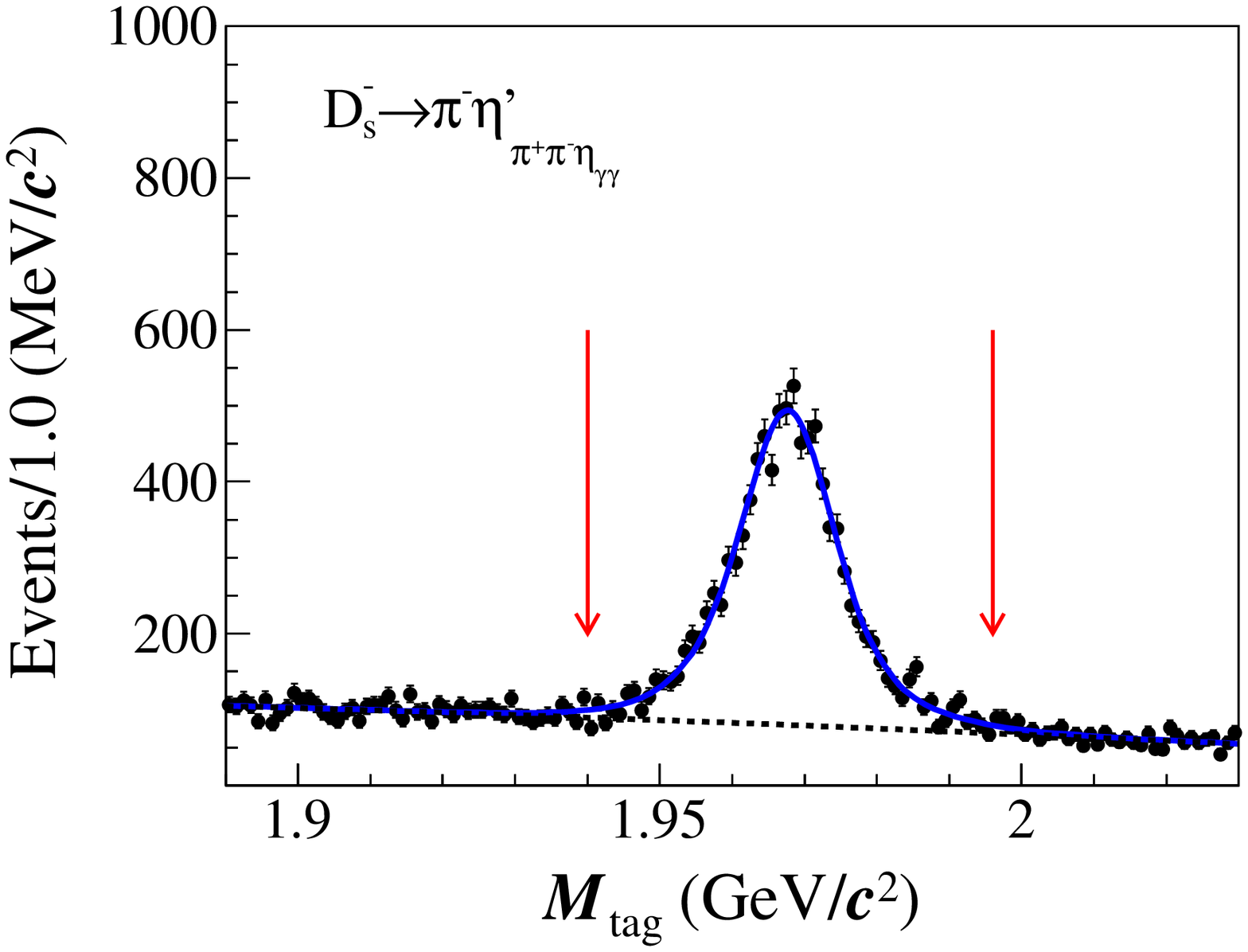}\\
  \includegraphics[width=0.4\textwidth]{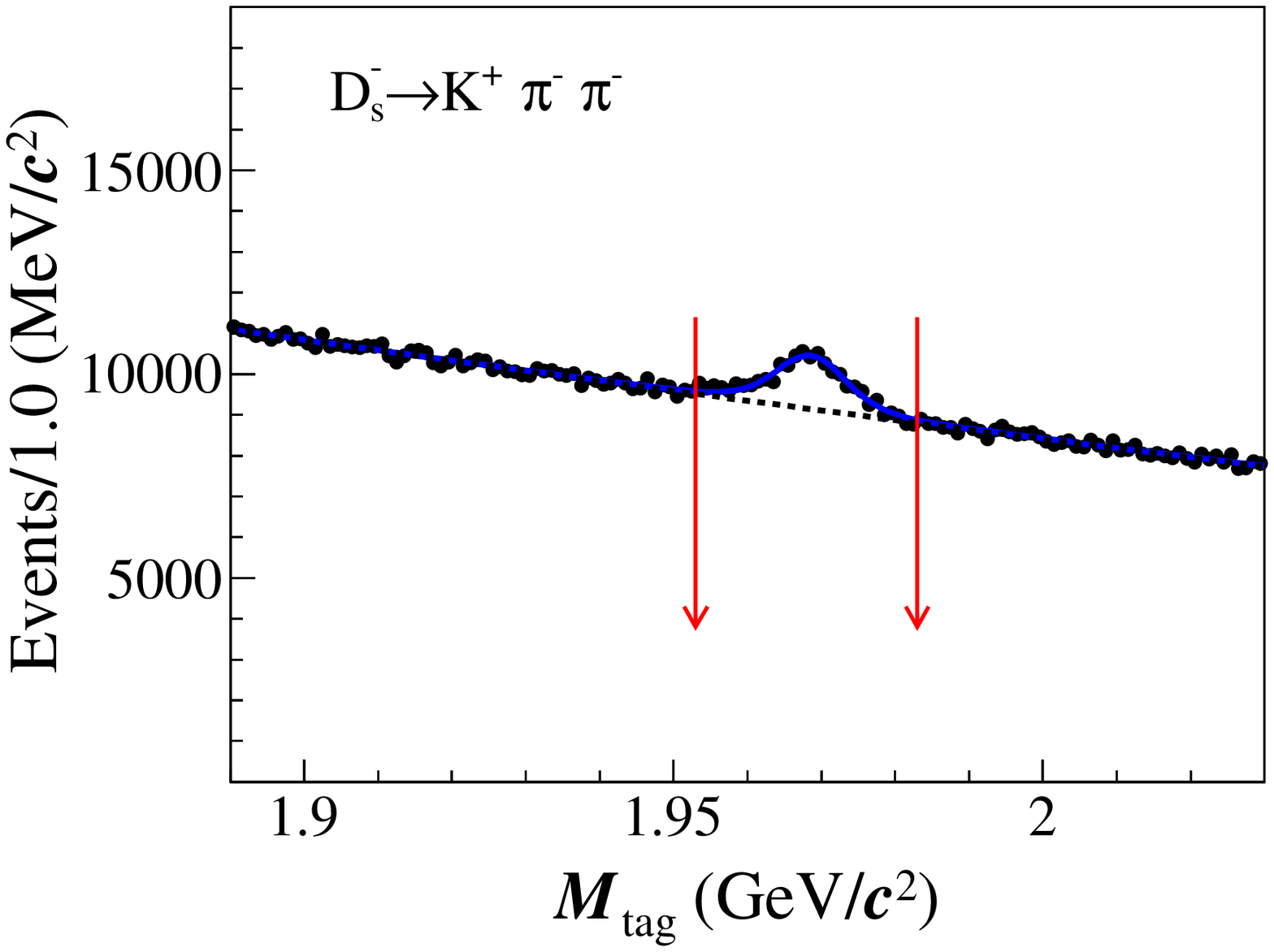}
\caption{Fits to the $M_{\rm tag}$ distributions of the ST candidates
         from the data sample at $\sqrt{s}=4.178$~GeV. The points with
         error bars are data, the blue solid lines are the total fits, and the black
         dashed lines are background. The pairs of red arrows denote the
         signal regions.
         }
\label{fit:Mass-data-Ds_4180}
\end{center}
\end{figure*}

The yields for various tag modes are obtained by fitting the corresponding
$M_{\rm tag}$ distributions and listed in Table~\ref{ST-eff}. As an example,
the fits to the $M_{\rm tag}$ spectra of the ST candidates in the data sample
at $\sqrt s=4.178$~GeV are shown in Fig.~\ref{fit:Mass-data-Ds_4180}. In the
fits, the signal is modeled by a MC-simulated shape convolved with a Gaussian
function to take into account the data-MC resolution difference. The background
is described by a second-order Chebyshev function. MC studies show that there is
no significant peaking background in any tag mode, except for
$D^{-} \to K_{S}^{0} \pi^-$ and $D_{s}^{-} \to \eta\pi^+\pi^-\pi^-$ faking the
$D_{s}^{-} \to K_{S}^{0} K^-$ and $D_{s}^{-} \to \pi^-\eta^{\prime}$ tags,
respectively. Therefore, the MC-simulated shapes of these two peaking background
sources are added to the background models.
\begin{table*}[htbp]
  \caption{The ST yields for the samples collected at $\sqrt{s} =$ (I) 4.178~GeV, (II) 4.199-4.219~GeV,
    and (III) 4.226~GeV. The uncertainties are statistical.}\label{ST-eff}
    \begin{center}
      \begin{tabular}{lccc}
        \hline
        Tag mode                                    & (I) $N_{\rm ST}$           & (II) $N_{\rm ST}$        & (III) $N_{\rm ST}$      \\
        \hline
        $D_{s}^{-}\to K_{S}^{0}K^{-}$               & $\phantom{0}31941\pm312\phantom{0}$   & $\phantom{0}18559\pm261$  & $\phantom{0}6582\pm160$ \\
        $D_{s}^{-}\to K^{+}K^{-}\pi^{-}$            & $137240\pm614\phantom{0}$             & $\phantom{0}81286\pm505$  & $28439\pm327$           \\
        $D_{s}^{-}\to K_{S}^{0}K^{-}\pi^{0}$        & $\phantom{0}11385\pm529\phantom{0}$   & $\phantom{00}6832\pm457$  & $\phantom{0}2227\pm220$ \\
        $D_{s}^{-}\to K_{S}^{0}K^{-}\pi^{-}\pi^{+}$ & $\phantom{00}8093\pm326\phantom{0}$   & $\phantom{00}5269\pm282$  & $\phantom{0}1662\pm217$ \\
        $D_{s}^{-}\to K_{S}^{0}K^{+}\pi^{-}\pi^{-}$ & $\phantom{0}15719\pm289\phantom{0}$   & $\phantom{00}8948\pm231$  & $\phantom{0}3263\pm172$ \\
        $D_{s}^{-}\to \pi^{-}\eta^{\prime}$         & $\phantom{00}7759\pm141\phantom{0}$   & $\phantom{00}4428\pm111$  & $\phantom{0}1648\pm74\phantom{0}$ \\
        $D_{s}^{-}\to K^{-}\pi^{+}\pi^{-}$          & $\phantom{0}17423\pm666\phantom{0}$   & $\phantom{0}10175\pm448$  & $\phantom{0}4984\pm458$ \\
        \hline
        Total                                       & $229560\pm1186$                       & $135497\pm937$            & $48805\pm688$ \\
        \hline
      \end{tabular}
    \end{center}
\end{table*}

Once a tag mode is identified, the signal decay
$D_{s}^{+} \to \pi^{+}\pi^{0}\pi^{0}$ is searched for at the recoiling side.
In the case of multiple candidates, the
DT candidate with the average mass, $(M_{\rm sig}+M_{\rm tag})/2$,
closest to the $D_{s}^{+}$ nominal mass is retained.
A $K^0_S\rightarrow \pi^0\pi^0$ mass veto,
$M_{\pi^0\pi^0}\notin (0.458, 0.520)$~GeV/$c^2$,
is applied on the signal $D_{s}^+$ to remove the
peaking background $D_s^{+}\to K_{S}^{0}\pi^+$.

To measure the BF, we start from the following equations for each tag mode:
\begin{eqnarray}\begin{aligned}
  N_{\text{tag}}^{\text{ST}} = 2N_{D_{s}^{*+}D_{s}^{-}}\mathcal{B}_{\text{tag}}\epsilon_{\text{tag}}^{\text{ST}}\,, \label{eq-ST}
\end{aligned}\end{eqnarray}
\begin{equation}
  N_{\text{tag,sig}}^{\text{DT}}=2N_{D_{s}^{*+}D_{s}^{-}}\mathcal{B}_{\text{tag}}\mathcal{B}_{\text{sig}}\epsilon_{\text{tag,sig}}^{\text{DT}}\,,
  \label{eq-DT}
\end{equation}
where $N_{D_{s}^{*+}D_{s}^{-}}$ is the total number of $D_{s}^{*\pm}D_{s}^{\mp}$
pairs produced from the $e^{+}e^{-}$ collisions; $N_{\text{tag}}^{\text{ST}}$ is
the ST yield for the tag mode; $N_{\text{tag,sig}}^{\text{DT}}$ is the DT yield;
$\mathcal{B}_{\text{tag}}$ and $\mathcal{B}_{\text{sig}}$ are the BFs of the
tag and signal modes, respectively; $\epsilon_{\text{tag}}^{\text{ST}}$ is the
ST efficiency to reconstruct the tag mode; and $\epsilon_{\text{tag,sig}}^{\text{DT}}$
is the DT efficiency to reconstruct both the tag and the signal decay modes. In
the case of more than one tag modes and sample groups,
\begin{eqnarray}
\begin{aligned}
  \begin{array}{lr}
    N_{\text{total}}^{\text{DT}}=\Sigma_{\alpha, i}N_{\alpha,\text{sig},i}^{\text{DT}}   = \mathcal{B}_{\text{sig}}
 \Sigma_{\alpha, i}2N_{D_{s}^{+}D_{s}^{-}}\mathcal{B}_{\alpha}\epsilon_{\alpha,\text{sig}, i}^{\text{DT}}\,,
  \end{array}
  \label{eq-DTtotal}
\end{aligned}
\end{eqnarray}
where $\alpha$ represents tag modes in the $i^{\rm th}$ sample group.
By isolating $\mathcal{B}_{\text{sig}}$, we find
\begin{eqnarray}\begin{aligned}
  \mathcal{B}_{\text{sig}} =
  \frac{N_{\text{total}}^{\text{DT}}}{ \mathcal{B}^2_{\pi^0\to\gamma\gamma}\begin{matrix}\sum_{\alpha, i} N_{\alpha, i}^{\text{ST}}\epsilon^{\text{DT}}_{\alpha,\text{sig},i}/\epsilon_{\alpha,i}^{\text{ST}}\end{matrix}}\,,
\end{aligned}\end{eqnarray}
where $N_{\alpha,i}^{\text{ST}}$ and $\epsilon_{\alpha,i}^{\text{ST}}$ are
obtained from the data and inclusive MC samples, respectively.
$\epsilon_{\alpha,\text{sig},i}^{\text{DT}}$ is determined with signal MC
samples with $D_{s}^{+} \to \pi^{+}\pi^{0}\pi^{0}$ events are generated
according to the results of the amplitude analysis.  The BF for $\pi^0\to\gamma\gamma$ 
is introduced to account for the
fact that the signal is reconstructed through this decay.

\begin{figure}[!htbp]
  \centering
  \includegraphics[width=0.6\textwidth]{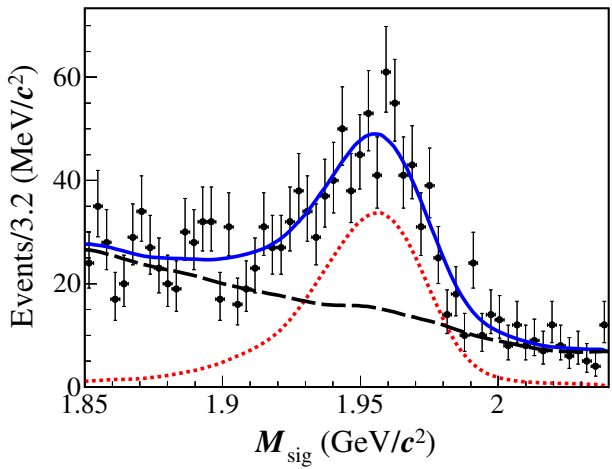}
 \caption{Fit to the $M_{\rm sig}$ distribution of the DT candidates from the
   data samples  at $\sqrt{s}= 4.178$-$4.226$~GeV. The data are represented by
   points with error bars, the total fit by the blue solid line, and the fitted
   signal and the fitted background by the red dotted and the black dashed
   lines, respectively.
 }
  \label{DT-fit}
\end{figure}

The DT yield $N_{\text{total}}^{\text{DT}}$ is found to be $587\pm44$ from the
fit to the $M_{\rm sig}$ distribution of the selected $D^+_s\to \pi^+\pi^0\pi^0$
candidates. The fit result is shown in Fig.~\ref{DT-fit}, where the signal
shape is described by a MC-simulated shape convolved with a Gaussian function
to take into account the data-MC resolution difference. The background is
described by a simulated shape from the inclusive MC sample. A small peaking background
originating from $D^0\to K^-\pi^+\pi^0$ is considered in the inclusive MC sample.
Taking the difference in $\pi^{0}$ reconstruction efficiencies for each signal mode
between data and MC simulation into account by multiplying the efficiencies by a factor of
99.5\% for each $\pi^{0}$, we determine the BF of $D_{s}^{+} \to \pi^{+}\pi^{0}\pi^{0}$
to be $(0.50\pm 0.04_{\rm stat}\pm 0.02_{\rm syst})\%$.

The relative systematic uncertainty for the total yield of the ST $D_s^-$ mesons
is assigned to be 0.4\% by examining the changes of the fit yields when varying
the signal shape, background shape, and taking into account the background
fluctuation in the fit. The systematic uncertainty due to the signal shape is
studied by repeating the fit without the convolved Gaussian. The MC-simulated
background shape is altered by varying the relative fractions of the dominant
backgrounds from $e^+e^-\to q\bar{q}$ or non-$D_{s}^{*+}D_{s}^{-}$ open-charm
processes by their statistical uncertainties of their related cross sections.
The largest change is taken as the corresponding systematic uncertainty. The
$\pi^{+}$ tracking (PID) efficiency is studied with the processes
$e^+e^-\to K^+K^-\pi^+\pi^-$ ($e^+e^-\to K^+K^-\pi^+\pi^-(\pi^0)$ and
$\pi^+\pi^-\pi^+\pi^-(\pi^0)$). The systematic uncertainty due to tracking (PID) efficiency
is estimated to be 1\%(1\%). The systematic uncertainty of the $\pi^{0}$
reconstruction efficiency is investigated by using a control sample of the
process $e^+e^-\to K^+K^-\pi^+\pi^-\pi^0$. The selection criteria listed in
Sec.~\ref{ST-selection} are used to reconstruct the two kaons and the two
pions. The recoiling mass distribution of $K^+K^-\pi^+\pi^-$ is fitted to
obtain the total number of $\pi^0$'s and the $\pi^0$ selection is applied to
determine the number of reconstructed $\pi^0$'s. The average ratio between
data and MC efficiencies of $\pi^0$ reconstruction, weighted by the
corresponding momentum spectra, is estimated to be $0.995 \pm 0.008$. After
correcting the simulated efficiencies to data by this ratio, the residual
uncertainty 0.8\% is assigned as the systematic uncertainty arising from each
$\pi^0$ reconstruction. The uncertainty due to the limited MC statistics is
obtained by
$\sqrt{\begin{matrix} \sum_{i} (f_{i}\frac{\delta_{\epsilon_{i}}}{\epsilon_{i}}\end{matrix}})^2$,
where $f_{i}$ is the tag yield fraction, and $\epsilon_{i}$ and
$\delta_{\epsilon_{i}}$ are the signal efficiency and the corresponding
uncertainty of tag mode $i$, respectively.
The uncertainty from the amplitude analysis model is estimated by varying the
model parameters based on their error matrix. The distribution of 600
efficiencies resulting from this variation is fitted by a Gaussian function and
the fitted width divided by the mean value is taken as a relative uncertainty. All of
the systematic uncertainties are summarized in Table~\ref{BF-Sys}. Adding them
in quadrature gives a total systematic uncertainty in the BF measurement of
4.0\%.
\begin{table}[htbp]
  \caption{Systematic uncertainties relative to the central value in the BF
    measurement.}
  \label{BF-Sys}
  \begin{center}
    \begin{tabular}{lccc}
      \hline
      Source   & Systematic uncertainty (\%)\\
      \hline
      $D_{s}^{-}$ yield                   & 0.4 \\
      Signal shape                        & 1.6 \\
      Background shape                    & 2.8 \\
      $\pi^{+}$ PID efficiency            & 1.0 \\
      $\pi^{+}$ tracking efficiency       & 1.0 \\
      $\pi^0$ reconstruction              & 1.6 \\
      MC statistics                       & 0.2 \\
      Signal MC model                     & 0.9 \\
      \hline
      Total                               & 4.0 \\
      \hline
    \end{tabular}
  \end{center}
\end{table}

\section{Summary}
An amplitude analysis of the decay $D_{s}^{+} \to \pi^{+}\pi^{0}\pi^{0}$ has
been performed for the first time. Amplitudes with significances larger
than $3\sigma$ were selected. The results for the FFs and phases of
the different intermediate processes are listed in Table~\ref{fit-result}.
With the detection efficiency calculated according to the intermediate processes found in the amplitude analysis, the BF for the decay
$D^+_s\to \pi^+\pi^0\pi^0$ is measured to be
$(0.50\pm 0.04_{\rm stat}\pm 0.02_{\rm syst})\%$. The precision is
improved by about a factor of two compared to the PDG value~\cite{PDG} due to the
large dataset collected with the BESIII detector. The BFs for the intermediate
processes are calculated with
$\mathcal{B}_{i} = {\rm FF}_{i} \times \mathcal{B}(D_{s}^{+} \to \pi^{+}\pi^{0}\pi^{0})$
and listed in Table~\ref{inter-processes}. The BF of $D^+_s\to f_0(980)\pi^+$
with $f_0(980)\to \pi^0\pi^0$ is measured for the first time. In addition, no significant
signal of $f_0(500)$ is observed. Assuming the BF ratio between $f_{0(2)}\to\pi^+\pi^-$ and $f_{0(2)}\to\pi^0\pi^0$ to be 2 based on isospin symmetry, 
our results favors with those from
$D^+_s\to \pi^+\pi^+\pi^-$ than from
$D_{s}^{+} \to K^{+}K^{-}\pi^{+}$.
\begin{table}[htbp]
  \caption{The BFs for intermediate processes.
    The first and the second uncertainties are
    statistical and systematic, respectively.}\label{inter-processes}
  \begin{center}
    \begin{tabular}{lc}
      \hline
      Intermediate process & BF ($10^{-3}$)\\
      \hline
	$D_{s}^{+} \to f_0(980)\pi^{+}$, $f_0(980)\to\pi^0\pi^0$   & $2.1\pm 0.3\pm 0.3 $  \\
	$D_{s}^{+} \to f_0(1370)\pi^{+}$, $f_0(1370)\to\pi^0\pi^0$  & $1.3\pm 0.2\pm 0.2 $  \\
	$D_{s}^{+} \to f_2(1270)\pi^{+}$, $f_2(1270)\to\pi^0\pi^0$  & $0.8\pm 0.3\pm 0.3 $  \\
	$D_{s}^{+} \to \pi^{+}(\pi^{0}\pi^{0})_{D}$   & $1.0\pm 0.3\pm 0.3 $  \\
	$D_{s}^{+} \to (\pi^{+}\pi^{0})_{D}\pi^{0}$   & $0.5\pm 0.2\pm 0.3 $  \\
      \hline
    \end{tabular}
  \end{center}
\end{table}

\acknowledgments
The BESIII collaboration thanks the staff of BEPCII and the IHEP computing center for their strong support. This work is supported in part by National Key R\&D Program of China under Contracts Nos. 2020YFA0406400, 2020YFA0406300; National Natural Science Foundation of China (NSFC) under Contracts Nos. 11625523, 11635010, 11735014, 11822506, 11835012, 11875054, 11935015, 11935016, 11935018, 11961141012, 12022510, 12025502, 12035009, 12035013, 12061131003; the Chinese Academy of Sciences (CAS) Large-Scale Scientific Facility Program; Joint Large-Scale Scientific Facility Funds of the NSFC and CAS under Contracts Nos. U2032104, U1732263, U1832207; CAS Key Research Program of Frontier Sciences under Contract No. QYZDJ-SSW-SLH040; 100 Talents Program of CAS; INPAC and Shanghai Key Laboratory for Particle Physics and Cosmology; ERC under Contract No. 758462; European Union Horizon 2020 research and innovation programme under Contract No. Marie Sklodowska-Curie grant agreement No 894790; German Research Foundation DFG under Contracts Nos. 443159800, Collaborative Research Center CRC 1044, FOR 2359, FOR 2359, GRK 214; Istituto Nazionale di Fisica Nucleare, Italy; Ministry of Development of Turkey under Contract No. DPT2006K-120470; National Science and Technology fund; Olle Engkvist Foundation under Contract No. 200-0605; STFC (United Kingdom); The Knut and Alice Wallenberg Foundation (Sweden) under Contract No. 2016.0157; The Royal Society, UK under Contracts Nos. DH140054, DH160214; The Swedish Research Council; U. S. Department of Energy under Contracts Nos. DE-FG02-05ER41374, DE-SC-0012069.
\bibliographystyle{JHEP}
\bibliography{references}

\end{document}